\documentclass[11pt]{article}

\usepackage[final]{acl}

\usepackage{times}
\usepackage{latexsym}
\usepackage{float}

\usepackage{algorithm}
\usepackage{algpseudocode}
\usepackage{amsmath}
\usepackage{amssymb}
\usepackage{rotating}
\usepackage[T1]{fontenc}

\usepackage[utf8]{inputenc}

\usepackage{microtype}

\usepackage{inconsolata}

\usepackage{graphicx}
\usepackage{inconsolata}
\usepackage{booktabs}
\usepackage{soul}
\usepackage{color}

\usepackage{multirow}
\usepackage{multicol}
\usepackage{enumitem,kantlipsum}
\usepackage[normalem]{ulem}

\usepackage{color,colortbl}
\usepackage{todonotes} 
\usepackage{soul}
\usepackage{threeparttable, makecell}
\usepackage{makecell}
\usepackage{xcolor}

\usepackage{comment}
\usepackage{array}  
\usepackage{lscape}  
\usepackage{listings} %
\usepackage{multirow}
\usepackage{graphicx}
\usepackage[normalem]{ulem}
\useunder{\uline}{\ul}{}

\usepackage{array}  
\usepackage{multirow}  
\usepackage{amsmath}
\usepackage{subcaption}
\usepackage{fancybox}
\usepackage{tikz}

\usepackage{arabtex}
\usepackage{utf8}
\setcode{utf8}

\definecolor{codegray}{gray}{0.9}

\lstdefinestyle{pythonstyle}{
    backgroundcolor=\color{codegray},   
    language=Python,
    basicstyle=\ttfamily\footnotesize,
    keywordstyle=\color{blue},
    stringstyle=\color{red},    
    breaklines=true,
    frame=single,
    keepspaces=true,
    showstringspaces=false,
}

\lstdefinestyle{aclprompt}{
  basicstyle=\ttfamily\small,
  columns=fullflexible,
  breaklines=true,
  breakatwhitespace=true,
  showstringspaces=false,
  keepspaces=true,
  frame=single,
  framerule=0.4pt,
  rulecolor=\color{black!25},
  escapeinside=``
  }

\lstdefinestyle{promptstyle}{
  basicstyle=\ttfamily\footnotesize,
  breaklines=true,
  breakatwhitespace=false,
  columns=fullflexible,
  keepspaces=true,
  showstringspaces=false,
  frame=single,
  framerule=0.3pt,
  rulecolor=\color{black!25},
  xleftmargin=0.5em,
  xrightmargin=0.5em,
  aboveskip=0.6em,
  belowskip=0.6em
}

\makeatletter
\renewcommand{\footnoterule}{%
  \kern-3pt
  \hrule width 0.4\columnwidth
  \kern 2.6pt
}
\makeatother
\usepackage{graphicx}
\definecolor{lightblue}{rgb}{.50,.95,1}
\definecolor{tri}{rgb}{.25,.88,.82}
\definecolor{lilac}{rgb}{0.85,0.64,0.85}

\newcommand{\msb}{\emph{MenaSpeechBank}}

\usepackage{ulem}
\usepackage{hyperref}
\usepackage{verbatim}
\usepackage{setspace}

\usepackage{colortbl}
\usepackage{booktabs}

\usepackage[edges]{forest}
\usepackage{forest}

\usetikzlibrary{arrows.meta}
\tikzset{>=Latex}

\title{\textit{MENASpeechBank:} A Reference Voice Bank with Persona-Conditioned Multi-Turn Conversations for SpeechLLMs}

\author{
Zien Sheikh Ali\thanks{~Equal contribution.},
Hunzalah Hassan Bhatti,
Rabindra Nath Nandi,\\
    {\bf
    Shammur Absar Chowdhury, 
Firoj Alam\textsuperscript{$*$}} \\
    Qatar Computing Research Institute, HBKU, Qatar\\ 
    \{fialam, shchowdhury\}@hbku.edu.qa  
}

\begin{document}
\maketitle
\begin{abstract}
\end{list}
Audio large language models (AudioLLMs) enable instruction following over speech and general audio, but progress is limited by the scarcity of \emph{diverse, conversational, and instruction-aligned} speech--text data. This gap is particularly pronounced for persona-grounded and dialectal interactions, where collecting real multi-speaker recordings remains costly and slow. We introduce \textsc{MENASpeechBank}, a reference speech bank comprising \textbf{\(\sim\)18K high-quality utterances} from \textbf{124 speakers} spanning multiple MENA countries, covering English, Modern Standard Arabic (MSA), and regional Arabic varieties. We develop a controllable data pipeline that \textit{(i)} constructs persona profiles enriched with World Values Survey (WVS) inspired attributes, \textit{(ii)} defines a taxonomy driven $\sim5K$conversational scenarios, \textit{(iii)} matches personas to scenarios via semantic similarity, \textit{(iv)} generates \(\sim\)417K role-play conversations with an LLM where the user speaks as the persona and the assistant behaves as a helpful agent, and \textit{(v)} produces speaker-conditioned user-turn audio (synthetic) from reference recordings to preserve speaker  diversity. We evaluate synthetic and human recorded conversations and provide an analysis. We will make the \textsc{MENASpeechBank} available for the community.\footnote{\href{https://huggingface.co/datasets/QCRI/MenaSpeechBank}{QCRI/MenaSpeechBank}} 
\end{abstract}

\section{Introduction}

\begin{figure}[t] 
    \centering
    \includegraphics[width=0.87\linewidth]{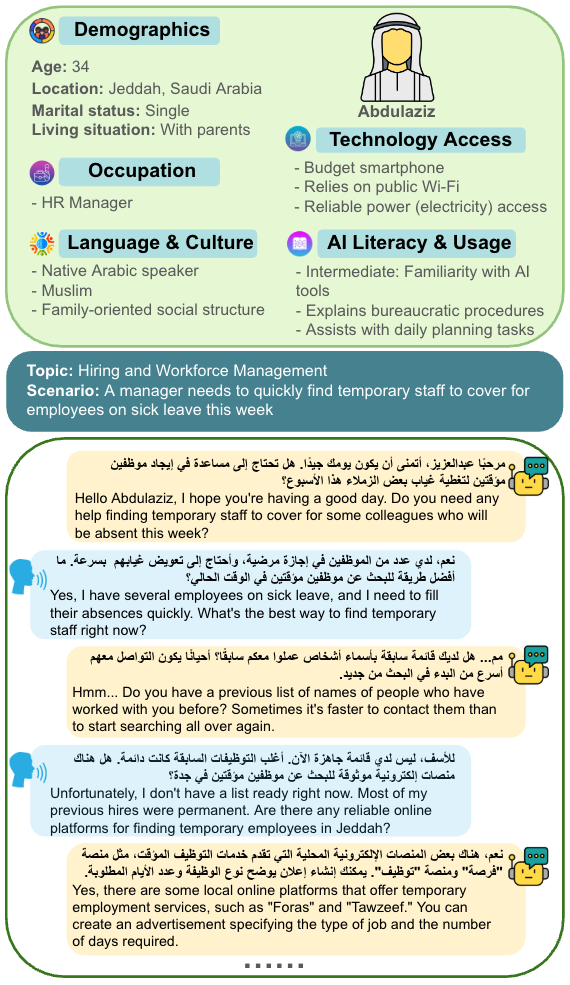} 
    \vspace{-0.2cm}
    \caption{An example of a persona, a conversational scenario, and conversational turns. Persona attributes are derived from basic demographic information, country-specific WVS values, and heuristics.}
    \label{fig:persona}
    \vspace{-0.5cm}
\end{figure}

Speech is the dominant interface for human communication, yet most large language models are still optimized for text. Recent %
Audio or Speech LLMs 
move beyond cascaded ``ASR $\rightarrow$ LLM'' systems by directly conditioning instruction following on audio inputs, enabling richer interaction patterns such as multi-turn spoken dialogue, speech-grounded reasoning, and paralinguistic understanding. Representative systems include Qwen-Audio~\cite{chu2023qwenaudio}, SpeechGPT~\cite{zhang-etal-2023-speechgpt}, AudioGPT~\cite{huang2024audiogpt}, and SALMONN~\cite{tang2024salmonn}. In parallel, evaluation suites such as AIR-Bench~\cite{yang-etal-2024-air} and AudioBench~\cite{wang-etal-2025-audiobench} increasingly emphasize open-ended, instruction-following behavior conditioned on diverse audio, surfacing persistent limitations in robustness, domain coverage, and conversational grounding.
A central bottleneck is \emph{data}. Compared to text-only instruction tuning, SpeechLLMs ~\cite{wang-etal-2025-benchmarking-contextual} require aligned audio--text pairs, and they benefit disproportionately from \emph{(i) conversational structure} (multi-turn context), \emph{(ii) task and intent diversity} (e.g., QA, classification, reasoning, dialogue acts), and \emph{(iii) speaker and dialect diversity} (accent, dialect, sociolect, bilingual/code-switched usage, and channel conditions). These desiderata are difficult to satisfy with real recordings alone: collecting persona-rich, long-form conversations is labor-intensive; obtaining broad dialect and speaker coverage is expensive and often infeasible for low-resource varieties; and licensing constraints frequently limit the release and reuse of speech data. This bottleneck is particularly pronounced in the Middle East and North Africa (MENA) region, where dialect diversity is high and 
resources remain limited~\cite{al2025landscape,ali_connecting_2021}.

Synthetic data pipelines are an attractive alternative, but their impact on SpeechLLMs is still under-explored. In text-only LLMs, self-generated instruction corpora can align models at scale (e.g., Self-Instruct~\cite{wang-etal-2023-self-instruct}). Persona conditioning has likewise been used to promote consistent dialogue (e.g., PersonaChat~\cite{zhang-etal-2018-personalizing}), and LLM-based generation with critics can improve persona faithfulness~\cite{jandaghi-etal-2024-faithful-persona}. In parallel, high-fidelity multi-speaker TTS and voice cloning now enable controllable, speaker-consistent synthesis~\cite{wang2025neuralcodec,le2023voicebox}, and combining LLM-generated text with TTS has begun to support speech-language instruction corpora~\cite{noroozi24_interspeech}. However, it remains unclear \emph{when} persona-grounded synthetic speech improves SpeechLLM performance, \emph{which} data factors drive gains (faithfulness, speaker/dialect diversity, synthesis quality, filtering), and \emph{how} to design end-to-end pipelines for downstream conversational and speech understanding tasks~\cite{chowdhury2016usersatis,stepanov2015automatic,ali2026wasilinthewildarabicspoken}.

We address this gap with \textsc{MENASpeechBank} (\msb{}), a data-centric study that pairs a \emph{multi-speaker MENA reference voice bank} with an \emph{end-to-end persona-to-speech generation pipeline} for SpeechLLM adaptation. Using speaker profiles as anchors, we construct controllable personas by completing attributes with World Values Survey (WVS) inspired values~\cite{haerpfer2020wvs7}, design a taxonomy driven conversational scenarios, and assign scenarios to personas via semantic similarity in an embedding space~\cite{reimers-gurevych-2019-sentence}. We then generate role-play dialogues with an LLM (persona as user; assistant as helpful agent), synthesize user turns from reference voices, and fine-tune an LLM on the resulting audio--text instruction pairs to evaluate gains on scenario-driven conversation and spoken QA. Our work makes the following key contributions:
\begin{itemize}[noitemsep,topsep=0pt,leftmargin=*,labelsep=.5em]
    \item We propose \msb{}, a multi-speaker MENA reference voice bank (\textbf{$\sim$18K utterances; 124 speakers}) covering \textbf{English and Arabic variants}.
    \item We enrich speaker profiles with WVS inspired value attributes to build controllable, behaviorally grounded personas.
    \item We propose an end-to-end persona $\rightarrow$ dialogue $\rightarrow$ speech pipeline using LLM role-play generation and speaker-conditioned voice cloning to produce scalable audio--text instruction data.
    \item We benchmark on synthetic and human recorded conversations.
\end{itemize}

\section{Related Work}
\label{sec:related work}

\noindent\textbf{Speech, Audio LLMs.}
A growing line of work equips LLMs with audio encoders and bridging modules to enable audio-conditioned instruction following. Examples include Qwen-Audio~\cite{chu2023qwenaudio}, SpeechGPT~\cite{zhang-etal-2023-speechgpt}, AudioGPT~\cite{huang2024audiogpt}, and SALMONN~\cite{tang2024salmonn}; related efforts such as Spectron~\cite{nachmani2024spectron} adapt LLMs for spoken question answering and speech generation using spectrogram-based representations. 
Benchmarking has evolved alongside these models. AIR-Bench~\cite{yang-etal-2024-air} and AudioBench~\cite{wang-etal-2025-audiobench} evaluate AudioLLMs under open-ended, chat-style instruction following across diverse audio types and task families, highlighting gaps in robustness and coverage and motivating data-centric improvements.

\noindent\textbf{Synthetic Instruction Data.}
Synthetic instruction tuning has been central to aligning text-only LLMs. Self-Instruct~\cite{wang-etal-2023-self-instruct} demonstrates that models can bootstrap large instruction corpora via iterative prompting and filtering. For dialogue, persona conditioning provides a lightweight mechanism for consistency and engagement. PersonaChat~\cite{zhang-etal-2018-personalizing} formalizes persona-grounded conversation and remains a standard testbed. More recently, LLMs have been used to generate persona-based conversational datasets at scale with explicit controls for adherence and faithfulness. \citet{jandaghi-etal-2024-faithful-persona} highlight both the benefits of synthetic persona data and the importance of careful filtering. \msb{} builds on these ideas, however, differs in two ways: \textit{(i)} we treat persona dialogue generation as \emph{scenario-conditioned} instruction generation for speech-centric evaluation, and \textit{(ii)} we render persona conversations into \emph{speech} with controllable speaker/dialect factors to study their effect on SpeechLLMs.

\noindent\textbf{Synthetic Speech and Training Data.}
Generative speech models now support high-quality multi-speaker TTS and voice cloning. Neural codec language models enable prompt-based voice cloning through discrete codec tokens~\cite{wang2025neuralcodec}, while Voicebox extends multilingual generation and speech editing~\cite{le2023voicebox}. Synthetic speech has also been used to create instruction data for adapting speech-language models with less reliance on labeled real audio~\cite{noroozi24_interspeech}. However, the role of \emph{persona-grounded conversational speech} in SpeechLLM adaptation remains underexplored, especially for SpokenQA and multi-turn dialogue where semantic understanding and speaker variation both matter. We study this setting through controlled speaker-conditioned data generation and conversational setups.

\section{MenaSpeechBank}
\label{sec:datasets}

\begin{figure}[t] 
    \centering
    \includegraphics[width=0.98\linewidth]{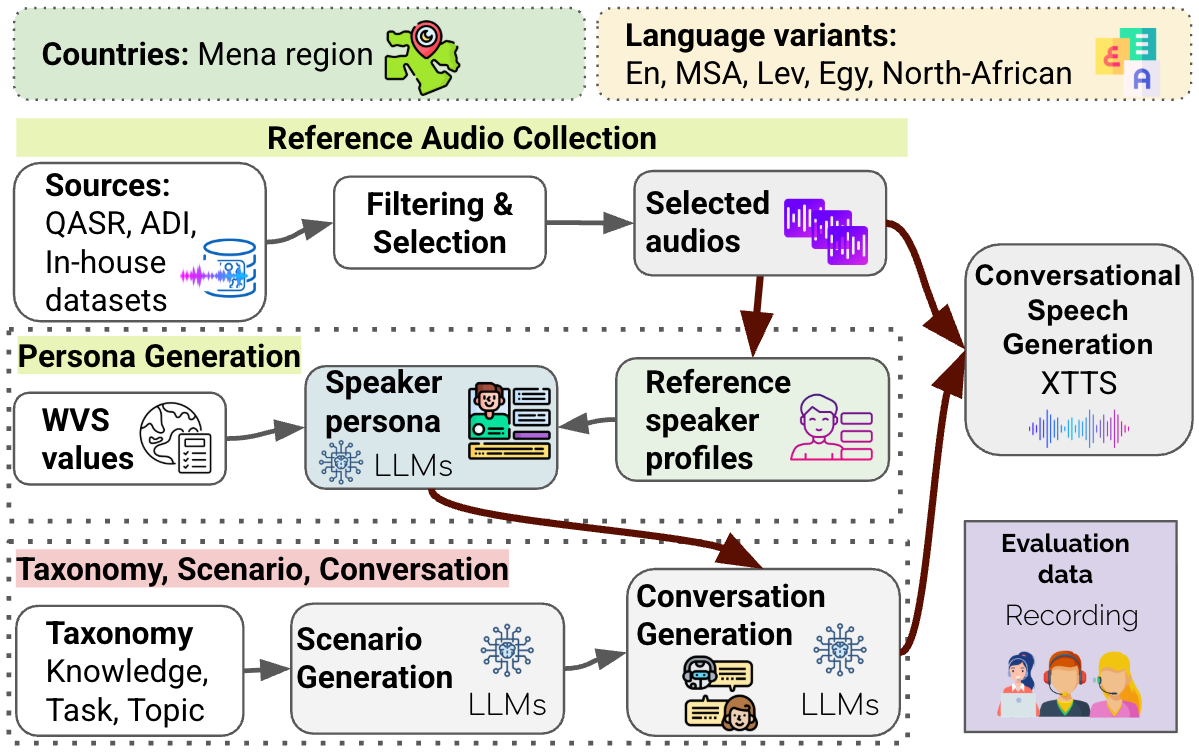} 
    \vspace{-0.3cm}
    \caption{An overview of \msb{} development pipeline.}
    \label{fig:speechbank_pipeline}
    \vspace{-0.3cm}
\end{figure}

In Figure~\ref{fig:speechbank_pipeline}, we present the \msb{} dataset development pipeline, which we describe in the following subsections.

\subsection{Reference Audio Collection}
We developed a reference speech bank with audio samples in MSA, dialectal arabic, and English for neural voice cloning based speech generation. 

\noindent
\textbf{Data Sources.} MSA reference audio was collected from different sources: QASR dataset \cite{mubarak-etal-2021-qasr}, and from \textit{in-house dataset}.
In addition to MSA, dialectal Arabic reference audio was collected from the \textit{in-house set}, and the ADI-17 dataset~\cite{shon2020adi17}, covering four major dialect regions: Egyptian, North African, Gulf, and Levantine Arabic. Finally, English reference audio was sourced from the English recordings available in the in-house dataset.

\noindent
\textbf{Filtering \& selection.} To ensure high audio quality and accurate text alignment, we applied dataset-specific filtering steps. For datasets with available reference scripts (our in-house dataset), we computed the word error rate (WER) between automatically generated transcriptions and the original text. Transcriptions were produced using the Fanar Transcription tool (Aura)\footnote{\url{https://api.fanar.qa/docs}} for Arabic speech and Whisper-Small\footnote{\url{https://huggingface.co/openai/whisper-small}} for English speech. We chose Fanar due to its free access and competitive performance compared to other state-of-the-art systems, as reported in \cite{alam2025spokennativqa}. We retained only samples with minimal transcription errors (WER $\leq$ 0.05). For the QASR dataset, we randomly selected 10 segments per speaker, each with a duration between 5 to 8 seconds, and used the automatic dialect identification tool Tamyiz \cite{abdullah2025voice} to identify samples classified as MSA, followed by manual annotation to verify that the selected samples were indeed in MSA. For the ADI-17 dataset, we manually inspected the selected segments and selected 56 high-quality samples from five speakers, ensuring that each sample contained speech from a single speaker. These steps ensure that the reference audios are clean, well-aligned with text, and consistent in language variety.

\subsection{Persona Generation}
\label{ssec_persona_generation}
For persona profile generation, we curated basic demographic information whenever it was available in the dataset. In addition, we used country-specific WVS values, simple heuristics, and LLMs. Below, we discuss these components in detail. 

\noindent
\textbf{Reference speaker profile.}
For the reference speaker profile, in some sources we had country, gender, age and educational information. Additionally for the paid speakers who recorded the audio and signed NDA for all permitted use of the data we asked them to provide basic demographic information. However, we do not ask or collect any personally identifiable information. Not for all speakers all of these information were available. For such cases we have derived them based on some heuristics (see Appendix~\ref{app:persona_generation_missing_data} for details). 

\noindent
\textbf{WVS value grounding.}
We have collected country specific WVS values (Wave 7)\footnote{World Values Survey (WVS) values capture population-level attitudes and beliefs about major aspects of social life, such as religiosity, family and gender norms, trust in others and institutions, political preferences and support for democracy, tolerance toward out-groups, work and economic views, and subjective well-being.} \cite{evs2022european} to include additional or missing attributes with the persona. We use country-level aggregates and cultural dimension summaries (e.g., traditional vs. secular-rational; survival vs. self-expression) to capture cross-country value differences. WVS value information is matched by matching speaker country and adjusted based on the age information.
More details on WVS incorporation are provided in Section \ref{app:persona_generation_wvs}.

\noindent
\textbf{Speaker persona profile.}
\label{speaker_profile1}
We construct persona profiles by combining \textit{(i)} basic metadata from reference speaker profiles, \textit{(ii)} WVS inspired value attributes, and \textit{(iii)} lightweight heuristics capturing technology access, AI literacy, and AI usage. Concretely, each persona is instantiated by sampling a country- and gender-consistent \emph{name} from curated name lists and a \emph{city} uniformly from a country-specific city inventory. We sample \emph{profession} uniformly from an age-bucketed occupation list, and for synthetic personas we draw \(\textit{age}\in[18,40]\) and \(\textit{gender}\in\{\text{male},\text{female}\}\), then sample \(\textit{marital status}\) and \(\textit{household type}\) uniformly from predefined categorical sets. To model digital and AI-related variation, we independently sample \emph{device} and \emph{connectivity} from fixed categorical inventories and assign an \emph{AI-competence level} from an ordinal set. \emph{AI use cases} are selected as a uniform sample of \(k{=}2\) distinct items from a predefined list. Finally, we generate a continuous OCEAN (Big Five) personality vector by perturbing fixed base values.

We use the original 124 speaker profiles as \emph{seeds} and expand them to create additional personas. To mitigate near-duplicates, we compute pairwise semantic similarity between persona descriptions using cosine similarity over multilingual Sentence-BERT embeddings~\cite{reimers-gurevych-2019-sentence}. We discard any candidate whose similarity to an existing profile exceeds 0.80, resulting in a final set of 469 personas.

\textbf{Persona summary.} Although personas are stored as structured attribute profiles, we convert each profile into a concise first-person \emph{persona summary} to better support LLM role-play generation and quality control. Persona-grounded dialogue work shows that conditioning on explicit persona descriptions improves consistency and personalization~\cite{li-etal-2016-persona,zhang-etal-2018-personalizing}. In our pipeline, the summary provides a compact interface for prompting and critic-based faithfulness checks~\cite{jandaghi-etal-2024-faithful-persona}, and a single text representation for embedding-based persona scenario matching~\cite{reimers-gurevych-2019-sentence}. We generate summaries with GPT-4.1 using  profile attributes; Appendix Figure~\ref{app_prompt_for_summary} lists the full prompt.

\textbf{Persona quality.} 
We automatically assess persona summaries using a deterministic, rule-based \emph{Persona Quality Index} (PQI) that counts satisfied checks over \textit{(i)} static attribute consistency with the persona profile (e.g., no unsupported numerals/proper nouns; no explicit mention of value terminology) and \textit{(ii)} narrative compliance (e.g., first-person voice, 90--180 words, grounded opening, single-paragraph prose, and a grounded ending). A summary is considered fully compliant only if it passes all checks (\(\mathrm{PQI}=12\)). 
On \(n=469\) personas, \(86.6\%\) achieve the maximum score, while the remaining cases are primarily attributable to a conservative \emph{consistency} flagging heuristic (\(11.1\%\)); full check definitions are provided in the Section \ref{ssec_app_persona_quality_assessment}.

\subsection{Taxonomy, Scenario and Conversations}

\noindent
\textbf{Taxonomy.}
For taxonomy development, we focused on \emph{everyday} user intents and information needs. We began with \textit{(i)} our predefined task/service domains (e.g., public sector \& utilities; travel \& mobility) and \textit{(ii)} knowledge-oriented topic categories drawn from \cite{alam2025everydaymmqa} and the everyday topics defined in \cite{hasan2024nativqa}. We harmonized overlapping labels, merged near-duplicates, and organized the result into a hierarchical taxonomy with two top-level branches: \textit{(i)} task/service domains and \textit{(ii)} knowledge/topic domains. Overall, the taxonomy is designed to reflect the \emph{service- and tool-mediated} interactions that dominate modern task-oriented conversational assistants, while maintaining broad coverage of major industries (e.g., finance, commerce, travel/mobility, health, public services, and productivity). The complete domain hierarchy is shown in Figure~\ref{fig:domain_taxonomy_root_left}. This design choice is consistent with established multi-domain dialogue benchmarks created to approximate real assistant behavior \citep{byrne2019taskmaster,rastogi2020sgd}.

\vspace{0.3cm}
\noindent
\textbf{Scenario.}
Although the taxonomy provides structured domain coverage, its leaf nodes remain too coarse to consistently elicit diverse, realistic conversations. Prompting directly from a leaf-domain label often produces repetitive or overly generic interactions, leaving \emph{common user intents} under-specified. We therefore introduce an intermediate \emph{topic} layer that operationalizes each root-to-leaf domain path. Specifically, for each domain path $d$, we use an LLM to generate $N=10$ topics (Listing~\ref{tab:app_topic_prompt}). We then generate scenarios for each domain path-topic pair $(d,t)$ as $\mathcal{S}_{d,t} \leftarrow \mathrm{LLM}_\theta(d,t;N)$, where $\mathcal{S}_{d,t}=\{s_i\}_{i=1}^{N}$ and $N=10$. In Listing \ref{tab_app_scenario_prompt}, we report the prompt. This intermediate expansion improves within-domain diversity and coverage prior to full conversation generation, consistent with observations in prior work \citep{sreedhar-etal-2024-canttalkaboutthis}. In this step we also applied near duplicate filtering following the same similarity based approach mentioned in Section \ref{speaker_profile1} with a threshold of 0.85. 
Using this procedure, we obtained 900 distinct topics and 4,521 scenarios.

\vspace{0.3cm}
\noindent
\textbf{Conversations.}
Because personas differ in roles, goals, and constraints, they are typically most plausible in different domain scenarios. Therefore, we align each persona with the scenario(s) that best match its summary. For each persona--scenario pair, we compute a hybrid similarity score that combines semantic similarity (cosine similarity of sentence-transformer embeddings; \texttt{sentence-transformers/all-MiniLM-L6-v2})\footnote{\href{https://huggingface.co/sentence-transformers/all-MiniLM-L6-v2}{all-MiniLM-L6-v2}} with lexical similarity (token-overlap Jaccard similarity), and retain only pairs with a score of at least $0.05$. This procedure yields scalable and reproducible persona--scenario pairings, ensuring that personas are grounded in an explicit context. Further details on the persona--scenario matching procedure are provided in Appendix \ref{persona_scenario_matching}.

After matching personas to scenarios, we generate persona-profiled assistant-user conversations conditioned on the selected persona attributes and scenario context. To streamline this process, we use GPT-4.1 with a structured prompt that instantiates the persona and scenario for each dialogue. The full prompt is provided in Listings~\ref{tab_app_dialogue_prompts_1} and~\ref{tab_app_dialogue_prompts_2} .

\subsection{Conversational Speech Generation}
Following persona--scenario--conditioned dialogue generation, we use neural voice closing approach to generate speech for the MSA conversations using the XTTS-v2 \cite{casanova24_interspeech}\footnote{\url{https://huggingface.co/coqui/XTTS-v2}} model. In this study, we focus on MSA because dialectal support in Arabic voice-cloning systems was limited at the time of our experiments; extending the pipeline to Arabic dialects is left for future work.
We generated audio for user turns only, while assistant turns are kept as text. For each conversation, a single reference audio corresponding to the persona's speaker profile is used. 
In total, we generate $\sim$2.1M audio samples across 416K conversations. Table \ref{tab:conv_stats} summarizes the corresponding conversation statistics. On average, each conversation contains 5.15 user turns and 5.84 assistant turns, with user and assistant utterances averaging 18.13 and 32.04 words per turn, respectively.

\noindent
\textbf{Quality assessment.} To assess audio quality, we randomly select 100 synthesized samples per speaker and evaluate them using standard objective metrics: \textit{(i)} Word Error Rate (WER) to measure transcription accuracy, computed using Fanar Aura; \textit{(ii)} Speaker Cosine Similarity (SpkCos)~\citep{ravanelli2021speechbrain} to assess speaker consistency, based on embeddings from the \texttt{spkrec-ecapa-voxceleb} model; and \textit{(iii)} NISQA~\citep{mittag2021nisqa}, which predicts overall perceptual speech quality.

The generated speech exhibits a moderate word error rate (WER = 10\%), while maintaining good perceptual quality (NISQA = 3.6) and moderate speaker consistency across conversations (SpkCos = 0.49).
To better interpret the SpkCos similarity score, we compare it agains two baselines using the same speaker encoder: same-speaker human-human similarty accross different recordings, as an empirical upper bound under our setup, and different-speaker similarity.
Our results show that the synthetic-human speaker similarity (0.49) is substantially higher than different-speaker similarity (0.12) and is closer to the natural same-speaker similarity (0.57), indicating that  generated audios reflect natural within-speaker variations and are separated from different-speaker pairs.

\subsection{Dataset Statistics}
\label{ssec_data_statistics}
\paragraph{Reference speakers.} \msb{} covers a wide range of speakers, comprising \textit{124 unique speakers} from 18 different countries. Tables \ref{tab:speaker_mother_tounge} and \ref{tab:speakers_age_group} offer a detailed breakdown of the number of speakers, utterances by mother tongue and conversation statistics. Of these, 56\% are male, and the dominant age group is 20-29 years, with 50 speakers in this range (see details in  Appendix~\ref{appendix_stats}).

\subsection{Speakers Statistics}\label{appendix:speaker_stats}

\begin{table}[t]
\centering
\scriptsize
\setlength{\tabcolsep}{3pt}
\begin{tabular}{lrr@{\quad}lrr}
\toprule
\textbf{Mother tongue} & \textbf{Spk.} & \textbf{Utt.} &
\textbf{Mother tongue} & \textbf{Spk.} & \textbf{Utt.} \\
\midrule
Arabic (Egypt)        & 31 & 6,440 & Arabic (Jordan)       & 5 & 690 \\
Arabic (Iraq)         & 11 & 1,861 & Urdu                  & 2 & 694 \\
Arabic (Morocco)      & 10 & 688   & Arabic (UAE)          & 2 & 26 \\
Arabic (Syria)        & 10 & 1,081 & Arabic (Libya)        & 2 & 1,549 \\
Arabic (Algeria)      & 8  & 788   & Arabic (Qatar)        & 3 & 227 \\
Arabic (Lebanon)      & 8  & 198   & Hindi                 & 2 & 1,377 \\
Arabic (Palestine)    & 7  & 111   & Arabic (Kuwait)       & 2 & 12 \\
Arabic (Saudi Arabia) & 6  & 333   & Arabic (Yemen)        & 1 & 10 \\
Arabic (Sudan)        & 6  & 959   & Arabic (Tunisia)      & 5  & 571 \\
\bottomrule
\end{tabular}
\vspace{-0.15cm}
\caption{Distribution of speakers and utterances in the \msb{} dataset by mother tongue.}
\label{tab:speaker_mother_tounge}
\vspace{-0.3cm}
\end{table}

In Table~\ref{tab:msb_stats}, we summarizes the number of speakers for each language or dialect, together with the average number of utterances per speaker and the average utterance duration. The speaker counts are reported \emph{per language variant}: many speakers contributed recordings in both Arabic and English. Consequently, the per-variant counts sum to 177 speaker entries, but these include overlaps across variants. In total, the dataset contains \textbf{124 unique speakers} as mentioned earlier. As MSA and English dominate the collection, they naturally include more speakers and a higher average number of utterances per speaker.

\begin{table}[t]
\centering
\small
\setlength{\tabcolsep}{4pt}
\scalebox{0.85}{%
\begin{tabular}{lrrrr}
\toprule
\textbf{Lang/Dial} &
\textbf{\#Spk} &
\textbf{\#Utt} &
\textbf{Utt/Spk} &
\textbf{Dur/Utt (s)} \\
\midrule
MSA & 101 & 10,777 & 133  & 6 \\
English & 29 & 5,548 & 192 & 6 \\
Gulf  & 15 & 287 & 17 & 5 \\
Egyption & 12 & 515 & 45 & 5 \\
North African & 10 & 168 & 17 & 4 \\
Levantine  & 10 & 346 & 35 & 5 \\
\midrule
\textbf{Total}  & 177 & \textbf{17,641} & - & - \\
\bottomrule
\end{tabular}
}
\vspace{-0.2cm}
\caption{Speaker and utterance statistics in the dataset by language or dialect. \#Spk denotes the number of speakers, \#Utt the number of utterances, Utt/Spk the average number of utterances per speaker, and Dur/Utt the average utterance duration (in seconds).
\vspace{-0.4cm}
}
\label{tab:msb_stats}
\end{table}

\begin{table}[t]
\centering
\setlength{\tabcolsep}{3.5pt}
\scalebox{0.70}{%
\begin{tabular}{@{}lrlr@{}}
\toprule
\textbf{Metric} & \textbf{Value} & \textbf{Metric} & \textbf{Value} \\
\midrule
Conversations        & 416{,}297   & Asst. turns        & 2{,}431{,}479 \\
User turns           & 2{,}144{,}048 & Asst. turns/conv.  & 5.84 \\
User turns/conv.     & 5.15        & Asst. words/turn   & 32.04 \\
User words/turn      & 18.13       & User duration/turn & 11.01s \\
\bottomrule
\end{tabular}
}
\vspace{-0.2cm}
\caption{Statistics of MSA conversations. Asst.: Assistant.}
\label{tab:conv_stats}
\vspace{-0.4cm}
\end{table}

\paragraph{Reference Speech Samples.}
\msb{} consists of \textbf{17,641 speech utterances} comprising $\approx$26.4 hours of speech. Figure~\ref{fig:dataset_utter_freq} illustrates the distribution of speech utterances by language and by dataset. The majority of utterances are in MSA and English, and notably, our in-house dataset comprises 98\% of the total corpus. 

\begin{figure}[t] 
    \centering
    \includegraphics[width=0.92\linewidth]{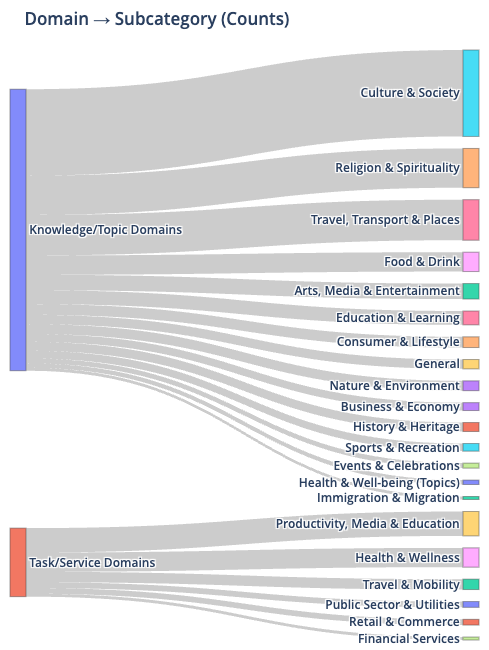} 
    \vspace{-0.23cm}
    \caption{Distribution of conversation domains and their respective subcategories.}
    \label{fig:domain_topic_catgegories}
    \vspace{-0.45cm}
\end{figure}

\paragraph{Domain and subcategory wise distribution.}
In Figure~\ref{fig:domain_topic_catgegories}, we summarize the distribution of conversations across domains and subcategories. A large portion of interactions fall under \textit{Knowledge/Topic} (80\%), alongside a substantial share of \textit{Task/Service} dialogues (20\%). This mix reflects the natural roles and intents of the reference speakers, which frequently elicit information-seeking and discussion-oriented scenarios while still capturing many actionable, service-oriented requests. Within \textit{Knowledge/Topic}, a small set of everyday themes contributes the most instances. \textit{Culture \& Society} accounts for 24.7\% of all samples (30.7\% within-domain), followed by \textit{Travel, Transport \& Places} (11.6\%) and \textit{Religion \& Spirituality} (11.2\%). These three subcategories comprise 47.5\% of the dataset. \textit{Task/Service} conversations are led by \textit{Productivity, Media \& Education} (7.0\% overall; 35.6\% within-domain) and \textit{Health \& Wellness} (5.5\% overall; 28.1\% within-domain), with \textit{Travel \& Mobility} adding a further 3.0\%. Overall, this distribution highlights strong coverage of culturally grounded interactions and tasks.

\section{Experimental Setup}
\label{sec:experiments}

\paragraph{Data splits.} 
For training and evaluation, we split the corpus \emph{by speaker profile} to ensure that conversations associated with the same underlying speaker/persona do not appear in multiple splits, thereby preventing overlap of speaker-grounded patterns (e.g., style, lexical preferences, or recurring personal facts) from train to test. We first reserve 12.3\% of the data as a test set at the profile level. The remaining 87.7\% is then partitioned into training and development sets in a 90/10 ratio using stratified sampling over scenarios, preserving the scenario distribution across splits. This procedure results in $\sim$328K
training, $\sim$36K %
development, and $\sim$51K %
test conversations.

\noindent\textbf{Models.} For baseline comparisons, we include strong closed-source models such as \textbf{Gemini-2.5 Pro} and \textbf{GPT-audio}, evaluated both in audio-native setting and within an ASR$\rightarrow$LLM pipeline. We also include open-weight models such as \textbf{Qwen} and Arabic-centric models including \textbf{Fanar-C-1-8.7B} \cite{fanar202422}, \textbf{Fanar-C-2-27B} \cite{fanarteam2026fanar20arabicgenerative}, and \textbf{Allam-7B} \cite{bari_allam}. %

\noindent\textbf{Fine-tuning.} 
We fine-tuned an in-house Qwen2.5-Omni-7B model that had been previously trained on 10K hours of ASR data. 
We fine-tuned this model on $\sim$330K multiturn conversations for 2 epochs. More details are provided in Appendix \ref{sec_app_fine_tuning}.  

\noindent\textbf{Evaluation data.} 
For benchmarking, we selected a subset of the \msb{} test split comprising 206 MSA conversations. Only the user turns were recorded, spoken by 10 paid Arabic-speaking annotators. Recording quality was verified using WER on ASR transcriptions generated by Fanar Aura, yielding an average WER of 12\%.

\noindent\textbf{Evaluation setup.}
We evaluate conversational response quality in an \emph{assistant-initiated} multi-turn setting where user inputs are \emph{audio} (synthetic and human-recorded). For each conversation, the model first receives the initial assistant and user-audio turns and produces a response. For subsequent turns, the model is conditioned on the  interaction history of prior user-assistant pairs (user audio and assistant text) to generate the next response. We report turn- and conversation-wise results.
Audio models are evaluated under two input conditions: \textit{(i)} direct audio input and \textit{(ii)} an ASR$\rightarrow$LLM cascaded pipeline, while text-only models are evaluated using the ASR$\rightarrow$LLM pipeline. Both settings are applied to speaker-conditioned synthetic and human-recorded conversations.

\noindent\textbf{LLM-as-a-Judge scoring.}
To score generated responses, we use an LLM-as-a-judge that evaluates both the turn level and conversation level, while using the full preceding transcript, the provided profile memory, and the session scenario as context. The judge outputs a structured checklist over ten rubrics: \textit{request understanding}, \textit{dialogue coherence}, \textit{context tracking}, \textit{profile memory consistency}, \textit{task completion}, \textit{constraint satisfaction}, \textit{specificity actionability}, \textit{calibration honesty}, \textit{language tone match} and \textit{cultural pragmatic fit}. We summarize quality using two metrics: \textit{Average Pass Rate (APR)}, the fraction of turns that jointly satisfy all required rubric checks, and \textit{Average Rubric Score (ARS)}, the mean pass rate across individual rubric checks. We use \textbf{\textit{APR as the primary metric}} in the main paper because it provides a stricter measure of overall response quality, while ARS is used for diagnostic analysis and reported in the Appendix \ref{sec_app_exp_results}. Similar rubric-based evaluation setup have been adopted in prior work~\cite{guo2025beyond,gosai2025audio}.

\section{Results and Discussion}
\label{sec:results}

Table~\ref{tab:apr_synth_human} reports conversation- and turn-level APR on speaker-conditioned synthetic and human-recorded inputs, %
while APR/ARS results, extended ablations, and fine-grained breakdowns are reported in Appendix \ref{sec_app_exp_results}.

\begin{table}[t]
\centering
\small
\setlength{\tabcolsep}{3.2pt}
\scalebox{0.92}{
\begin{tabular}{lcccc}
\toprule
&
\multicolumn{2}{c}{\textbf{Conversation}} &
\multicolumn{2}{c}{\textbf{Turn}} \\
\cmidrule(lr){2-3}\cmidrule(lr){4-5}
\textbf{Model}
& \textbf{Synth} & \textbf{Human}
& \textbf{Synth} & \textbf{Human} \\
\midrule

\multicolumn{5}{c}{\textbf{Large Closed Models}} \\
\midrule
Gemini-Audio              & 0.777 & 0.806 & 0.833 & 0.871 \\
Gemini (ASR)+ Gemini       & 0.828 & 0.815 & 0.877 & 0.853 \\
GPT-audio                   & 0.767 & 0.728 & 0.801 & 0.821 \\
GPT (ASR)+GPT-5.2     & 0.845 & 0.816 & 0.888 & 0.866 \\
\midrule

\multicolumn{5}{c}{\textbf{Open and Fine-tuned Models}} \\
\midrule
Qwen2.5-Omni-7B             & 0.024 & 0.034 & 0.256 & 0.226 \\
Qwen2.5-Omni-7B FT          & 0.252 & 0.204 & 0.559 & 0.534 \\
\midrule

\multicolumn{5}{c}{\textbf{Arabic-centric Cascaded Models}} \\
\midrule
Fanar(Aura)+Allam-7B & 0.262 & 0.248 & 0.562 & 0.508 \\
Fanar(Aura)+Fanar & 0.366 & 0.353 & 0.616 & 0.632 \\
Fanar(Aura)+Fanar-2  & 0.434 & 0.371 & 0.669 & 0.646 \\

\bottomrule
\end{tabular}%
}
\vspace{-0.2cm}
\caption{Conversation and turn level APR results on synthetic and human-recorded inputs. 
Full APR/ARS results are provided in Table~\ref{tab:full_results_appendix}.} 
\label{tab:apr_synth_human}
\vspace{-0.4cm}
\end{table}

\begin{table}[t]
\centering
\scalebox{0.83}{
\begin{tabular}{lcccc}
\toprule 
&
\multicolumn{2}{c}{\textbf{Conv.-level}} &
\multicolumn{2}{c}{\textbf{Turn-level}} \\
\textbf{Models}
& \textbf{APR} & \textbf{ARS} & \textbf{APR} & \textbf{ARS} \\
\midrule
Allam & 0.325 & 0.659 & 0.585 & 0.898 \\
Fanar-C-1-8.7B & 0.502 & 0.789 & 0.698 & 0.941  \\
Fanar-C-2-27B & 0.410 & 0.752 & 0.652 & 0.911 \\
\bottomrule
\end{tabular}
}
\vspace{-0.25cm}
\caption{Conversation- and turn-level results across text-only inputs for Arabic-centric models.}
\label{tab:text_only_arabic_models}
\vspace{-0.35cm}
\end{table}

\begin{table*}[t]
\centering
\small
\setlength{\tabcolsep}{4pt}
\renewcommand{\arraystretch}{1.1}
\scalebox{0.98}{
\begin{tabular}{llcccccccc}
\toprule
\multirow{2}{*}{\textbf{Model}} &
\multirow{2}{*}{\textbf{Input}} &
\multicolumn{4}{c}{\textbf{Conversation-level}} &
\multicolumn{4}{c}{\textbf{Turn-level}} \\
\cmidrule(lr){3-6}
\cmidrule(lr){7-10}
& &
\multicolumn{2}{c}{\textbf{Synthetic}} &
\multicolumn{2}{c}{\textbf{Human}} &
\multicolumn{2}{c}{\textbf{Synthetic}} &
\multicolumn{2}{c}{\textbf{Human}} \\
\cmidrule(lr){3-4}
\cmidrule(lr){5-6}
\cmidrule(lr){7-8}
\cmidrule(lr){9-10}
& &
\textbf{APR} & \textbf{ARS} &
\textbf{APR} & \textbf{ARS} &
\textbf{APR} & \textbf{ARS} &
\textbf{APR} & \textbf{ARS} \\
\midrule

\multicolumn{10}{c}{\textbf{Large Closed Models}} \\
\midrule

\multirow{3}{*}{Gemini-2.5 pro}
& Audio
& 0.777 & 0.892
& 0.806 & 0.926
& 0.833 & 0.971
& 0.871 & 0.980 \\

& Transcription (Fanar)
& 0.820 & 0.925
& 0.850 & 0.937
& 0.850 & 0.976
& 0.859 & 0.979 \\

& Transcription (Gemini)
& 0.828 & 0.937
& 0.815 & 0.925
& 0.877 & 0.979
& 0.853 & 0.978 \\

GPT-audio
& Audio
& 0.767 & 0.893
& 0.728 & 0.894
& 0.801 & 0.964
& 0.821 & 0.965 \\

GPT-5.2
& Transcription (GPT-audio)
& 0.845 & 0.947
& 0.816 & 0.947
& 0.888 & 0.979
& 0.866 & 0.977 \\

\midrule
\multicolumn{10}{c}{\textbf{Open \& FT Models}} \\
\midrule

\multirow{2}{*}{Qwen2.5-Omni-7B}
& Audio
& 0.024 & 0.314
& 0.034 & 0.271
& 0.256 & 0.734
& 0.226 & 0.712 \\

& Transcription (Fanar)
& 0.010 & 0.266
& 0.024 & 0.267
& 0.234 & 0.716
& 0.228 & 0.709 \\

\multirow{2}{*}{Qwen2.5-Omni-7B FT}
& Audio
& 0.252 & 0.624
& 0.204 & 0.597
& 0.559 & 0.887
& 0.534 & 0.871 \\

& Transcription (Fanar)
& 0.228 & 0.616
& 0.218 & 0.589
& 0.523 & 0.865
& 0.528 & 0.860 \\

\midrule
\multicolumn{10}{c}{\textbf{Arabic-centric Models}} \\
\midrule
Allam
& Transcription (Fanar)
& 0.262 & 0.631
& 0.248 & 0.587
& 0.562 & 0.886
& 0.508 & 0.871 \\
Fanar-C-1-8.7B
& Transcription (Fanar)
& 0.366 & 0.708
& 0.353 & 0.727
& 0.616 & 0.894
& 0.632 & 0.907 \\
Fanar-C-2-27B
& Transcription (Fanar)
& 0.434 & 0.759
& 0.371 & 0.734
& 0.669 & 0.923
& 0.646 & 0.921 \\

\bottomrule
\end{tabular}
}
\vspace{-0.2cm}
\caption{Conversation- and turn-level evaluation results across synthetic and human conversations for audio and transcription-based inputs.}
\label{tab:full_results_appendix}
\vspace{-0.3cm}
\end{table*}

\noindent
\textbf{Overall performance.}
Closed models provide the strongest reference results. GPT (ASR)+GPT-5.2 achieves the highest conversation-level APR on both synthetic and human inputs (0.845 and 0.816), while Gemini (ASR)+Gemini-2.5-Pro performs best at the turn level on synthetic recordings (0.877). These results indicate that both direct audio modeling and cascaded ASR$\rightarrow$LLM pipeline remain competitive for this task, with the cascaded system benefiting from stronger text-side reasoning after transcription.

\noindent
\textbf{Turn-level vs. conversation-level quality.}
Across all systems, turn-level APR is higher than conversation-level APR. This is expected because conversation-level evaluation is stricter and reflects accumulated failures across multiple turns. The gap is especially large for weaker open-weight models, indicating that isolated turns may pass some requirements, while maintaining speakers' profile-aware, and scenario-consistent behavior over a \textit{full conversation remains challenging}.

\noindent
\textbf{Does performance differ between synthetic and human speech inputs?}
Human recordings do not uniformly degrade performance, but they introduce greater variability across systems. For example, Gemini-2.5 Pro improves from 0.777 to 0.806 at the conversation level and from 0.833 to 0.871 at the turn level, suggesting robust handling of human-recorded speech. In contrast, GPT-audio and GPT (ASR)+GPT-5.2 show lower conversation-level APR on human inputs. This pattern suggests that \textit{human speech affects systems differently depending on their audio understanding and transcription quality}.

\noindent
\textbf{Does fine-tuning help?}
Fine-tuning substantially improves the open-weight Qwen2.5-Omni-7B model. At the conversation level, APR improves from 0.024 to 0.252 on synthetic inputs and from 0.034 to 0.204 on human recordings. At the turn level, the gains are even larger, increasing from 0.256 to 0.559 on synthetic inputs and from 0.226 to 0.534 on human recordings. These results show that task-specific \textit{fine-tuning improves both turn and conversational responses}.

\noindent
\textbf{Arabic-centric cascaded models.}
Since no end-to-end AudioLLM is specifically designed for Arabic conversational speech in our setting, we evaluate Arabic-centric ASR$\rightarrow$LLM pipelines using Fanar-Aura ASR followed by ALLaM and Fanar models. Among these systems, Fanar(Aura)+Fanar-C-2-27B performs best, reaching 0.434 conversation-level APR on synthetic inputs and 0.371 on human recordings, with turn-level APR of 0.669 and 0.646, respectively. Across all four evaluation settings, Fanar-C-2-27B outperforms both ALLaM and Fanar-C-1-8.7B when paired with the same Fanar-Aura ASR frontend. This suggests that stronger Arabic-centric text reasoning improves the cascaded pipeline after ASR. However, the remaining gap to strong closed-model performance highlights \textit{the need for Arabic speech models}.

\noindent
\textbf{ARS as diagnostic evidence.}
While APR is our primary metric, ARS provides useful diagnostic information by measuring average rubric-level satisfaction. In Appendix Table~\ref{tab:full_results_appendix}, we observe cases where ARS is substantially higher than APR, indicating that models often satisfy several rubric dimensions but fail at least one required criterion. This supports our choice of APR as the main metric, since it better captures complete response success in multi-turn conversational evaluation.

\noindent
\textbf{LLM-judge agreement with humans.}
To assess evaluation reliability, we manually annotated the outputs of 50 Qwen2.5-Omni-7B (Base) conversations and 50 Qwen2.5-Omni-7B FT (2 epochs) conversations using the same rubric and checklist used by our LLM judge. The fine-tuned model achieves 71\% observed agreement and a Gwet's AC1 \cite{gwet2008computing} of 0.52 with human judges, compared to 62\% and 0.25 for the base model. This indicates that the fine-tuned model provides better alignment with human judgments and supports its use as a scalable complement to human evaluation.

\noindent
\textbf{Scope of the evaluation.}
We evaluate a focused benchmark subset to establish initial baselines under compute, latency, and API-cost constraints. The subset retains key properties of the benchmark, including multi-turn interactions with both speaker-conditioned synthetic and human-recorded inputs, enabling  comparison across direct-audio, cascaded, closed, open-weight, and fine-tuned systems.

\section{\msb{}: Applications}
\msb{} is, to the best of our knowledge, \textbf{the first} \emph{bilingual} (Arabic-English) and \emph{MENA-centric} resource that jointly provides a reference speech bank, profile-conditioned multi-turn conversations, and corresponding speech. This combination enables controlled studies of spoken conversation while remaining close to real voice-assistant interactions. Below, we outline several use cases.

\noindent\textbf{Audio-to-text multi-turn dialogue benchmarking.}
\msb{} supports turn-level evaluation of AudioLLMs, measuring response quality, coherence, and context tracking when user inputs are provided directly as speech rather than transcripts~\cite{ahmed2026multiturn}. It enables evaluation of how well AudioLLMs jointly handle speech understanding and response generation across turns, including robustness to speaker variability, accents, and earlier conversational information.

\noindent\textbf{Long-context spoken dialogue memory evaluation.}
The multi-turn structure enables stress-testing of long-horizon conversational memory: whether models retain and correctly reuse constraints, preferences, and entities introduced in earlier \emph{spoken} turns, and whether they avoid hallucinating user-specific facts beyond the provided profile memory.

\noindent\textbf{Speech-conditioned personalization modeling.}
As each dialogue is grounded in an explicit persona profile, \msb{} can be used to study personalization under speech input, including consistent use of profile attributes, stable interaction style, and appropriate register across turns.

\noindent\textbf{Robustness and invariance probing under speech variability.}
By pairing controlled scenarios with diverse speakers and recordings, \msb{} facilitates analyses of robustness to dialect/accent variation, speaker identity, speaking rate, and recording conditions, while holding the underlying semantic intent fixed.

\noindent\textbf{Multi-turn conversational alignment for AudioLLMs.}
Finally, \msb{} can support preference-based optimization (e.g., reranking, rejection sampling, or preference tuning) using rubric- and pairwise-judged outputs to explicitly target multi-turn dialogue qualities such as coherence, calibration, and profile-consistent context tracking, rather than single-turn helpfulness alone.

\section{Conclusions}
\label{sec:conclusions}
We introduced \msb{}, to our knowledge \textbf{the first} bilingual Arabic-English MENA-centric speech bank and speaker-conditioned multi-turn conversation resource for evaluating and training speech-capable conversational models. We also presented an LLM-as-judge evaluation protocol with APR as the primary metric and ARS as a diagnostic measure, covering relevance, completeness, specificity, coherence, context tracking, calibration, language, and safety. Our experiments show that strong closed-model perform best, while fine-tuning substantially improves open-weight SpeechLLMs and Arabic-centric ASR$\rightarrow$LLM pipelines provide competitive cascaded baselines. We hope \msb{} will facilitate research on multi-turn spoken interaction, long-context memory, personalization, and robustness across speakers and recording conditions. Future work will scale evaluation to the full benchmark and explore training strategies to improve persona consistency, and robustness to human speech.

\section{Limitations}
\label{sec:limitations}
\msb{} is intended as a controlled resource for studying persona-conditioned and speaker-conditioned conversational speech, rather than an exhaustive representation of all MENA populations, dialects, or speaking styles. Coverage across countries and dialect varieties can be further expanded and balanced, especially as more diverse reference speakers become available. The current persona schema also simplifies some sociolinguistic attributes through bounded age ranges and categorical fields. Similarly, the WVS-inspired attributes and heuristics for technology access and AI use provide controllability, but should be interpreted as modeling choices rather than demographic claims. Future work can refine these attributes to better capture social and linguistic nuance.

The generated conversations and speaker-conditioned audio may differ from naturally occurring dialogue in spontaneity, disfluency, turn-taking, prosody, and recording conditions. Although we use deterministic PQI checks and conservative checks to reduce common artifacts.

Additionally, the limited availability of high-quality conversational Arabic TTS models supporting multiple dialects at the time of this study led us to generate conversations primarily in MSA. Extending the benchmark to dialectal Arabic is left as future work, requiring substantial human effort to ensure natural, accurate, and dialect-representative conversations. 

Our evaluation includes both speaker-conditioned and human-recorded inputs, and it is conducted on a focused subset of the benchmark and emphasizes scenario-driven conversation and spoken QA.

\section*{Ethics Statement}
\label{sec:ethics}
Our dataset is designed to minimize privacy risk: we do not collect or store personally identifiable information, and the released artifacts are intended for research use under clear documentation. We use only high-level demographic cues from reference speakers as seeds and extend them with WVS-inspired attributes and heuristic signals, so the resulting persona profiles are synthetic and are not intended to correspond to any real individual. While synthetic personas enable controllable generation at scale, they can still surface representational biases or stereotypes inherited from the underlying schema, value priors, or scenario design. To mitigate this, our pipeline conditions summaries strictly on explicit profile fields and applies deterministic checks for groundedness and constraint compliance; we additionally recommend routine checks for representational harms (e.g., gender- or nationality-linked stereotypes) before release or deployment, and we position persona summaries as conditioning signals for data generation rather than as faithful depictions of real people.

\section*{Broader Impact}
\msb{} enables controllable, large-scale synthetic speech instruction data that can accelerate AudioLLM adaptation for underrepresented languages and speech varieties, supporting more inclusive spoken assistants, educational applications, and improved information access for MENA speakers while enabling systematic robustness studies across accents, dialects, and bilingual usage. To ensure these benefits are realized responsibly, we promote clear documentation and intended-use guidance, appropriate licensing and access controls, and routine check of personas and scenarios to reduce representational bias and discourage misuse (e.g., impersonation or misinformation), and we encourage future work to broaden evaluations across demographic groups and dialects alongside targeted risk assessments before deployment.

\bibliography{bibliography/bibliography}

\appendix

\label{sec:appendix}
\section{Details on Persona Generation}\label{app:persona_generation}

\subsection{Handling Missing Persona Data}
\label{app:persona_generation_missing_data}

If iso3 or country name is missing, we derive it from the normalized country name (and vice‑versa). We then sample culturally aligned name and city from country-specific lists; if unavailable, we fall back to regional/default lists. Profession is sampled from age-bucketed occupation pools, which are optionally constrained by education level. Household/digital-access fields (device/connectivity/AI competence) and 1–2 AI use cases are sampled from small categorical pools conditioned on age/education to ensure plausibility.

Figure \ref{app_persona_example} shows an example illustrating the input persona attributes and the corresponding derived outputs. We omit the descriptive persona summary for brevity.

\begin{figure}[t]
\centering
\begin{minipage}{\columnwidth}

\noindent\textbf{Speaker attributes.}
\begin{lstlisting}[
  style=promptstyle,
  basicstyle=\ttfamily\scriptsize,
  breaklines=true,
  breakatwhitespace=false,
  columns=fullflexible
]
{
  "speaker_id": "QASR_9b101a25d6ef3a5f300ac4c8fce2058f",
  "nationality": "Algeria",
  "age": 52,
  "gender": "Female",
  "mother_tongue": "Arabic",
  "education_level": "bachelor",
  "languages": ["msa"],
  "iso3": "DZA"
}
\end{lstlisting}

\vspace{-0.3em}
\noindent\textbf{Generated persona profile (excerpt).}
\begin{lstlisting}[
  style=promptstyle,
  basicstyle=\ttfamily\scriptsize,
  breaklines=true,
  breakatwhitespace=false,
  columns=fullflexible
]
{
  "speaker_id": "QASR_9b101a25d6ef3a5f300ac4c8fce2058f",
  "speaker_age": 52,
  "speaker_nationality": "Algeria",
  "speaker_mother_tongue": "Arabic",
  "native": true,
  "persona_name": "Samira",
  "gender": "female",
  "city": "Ain Oulmene",
  "education_level": "bachelor",
  "profession": "chief pharmacist",
  "marital_status": "married",
  "household_type": "lives with family",
  "digital_access": {
    "device": "smartphone and laptop",
    "connectivity": "public Wi-Fi occasionally",
    "ai_competence_level": "advanced"
  },
  "ai_use_cases": [
    "writing or editing professional emails",
    "trip or visa document checklist help"
  ],
  "wvs_profile": {
    "axis_rel_trad": 0.899,
    "axis_self_expr": 0.230,
    "axis_trust": 0.215,
    "axis_gender_eq": 0.657,
    "axis_inst_trust": 0.520,
    "axis_auth": 0.722,
    "axis_econ_egal": 0.719,
    "axis_env": 0.426,
    "axis_civic": 0.226
  },
  "ocean": {
    "openness": 0.571,
    "conscientiousness": 0.855,
    "extraversion": 0.547,
    "agreeableness": 0.621,
    "neuroticism": 0.309
  }
}
\end{lstlisting}

\end{minipage}
\vspace{-0.2cm}
\caption{Example speaker attributes and generated persona profile.}
\label{app_persona_example}
\end{figure}

\subsection{WVS Incorporation}\label{app:persona_generation_wvs}
We retrieve a country-level WVS vector (via ISO3/country name) as normalized value axes and use it as a soft conditioning prior that guides conversational framing (e.g., trust/tradition/self-expression) without asserting individual beliefs. If WVS is unavailable, we use a default axis vector with a small age-based adjustment.

\subsection{Persona Quality Assessment (PQI).}
\label{ssec_app_persona_quality_assessment}
To verify that each generated persona summary is (a) grounded in the underlying attribute record and (b) compliant with the required narrative constraints, we define a deterministic, rule-based \emph{Persona Quality Index} (PQI). PQI decomposes quality into two components: \emph{static fidelity} (\(\mathrm{PQI}_S\)), which measures whether the summary stays within the information supported by the persona attributes, and \emph{narrative compliance} (\(\mathrm{PQI}_N\)), which measures adherence to format and style constraints (e.g., first-person voice, grounded opening, length bounds, and a non-moralizing ending). For each persona \(i\), we compute a set of binary checks \(\{S_{i,j}\}_{j=1}^{J}\) (static) and \(\{N_{i,k}\}_{k=1}^{K}\) (narrative), where each indicator equals \(1\) if the corresponding requirement is satisfied and \(0\) otherwise. We then define:

\begin{equation}
\begin{alignedat}{1}
\mathrm{PQI}_S(i) &= \sum_{j=1}^{J} S_{i,j}, \\
\mathrm{PQI}_N(i) &= \sum_{k=1}^{K} N_{i,k}, \\
\mathrm{PQI}(i)   &= \mathrm{PQI}_S(i)+\mathrm{PQI}_N(i).
\end{alignedat}
\end{equation}

Here, \(j\) and \(k\) index the individual static and narrative checks, respectively, and \(J\) and \(K\) denote the total number of checks in each group.

In our implementation, \(\mathrm{PQI}_S\) comprises \(J{=}5\) grounding checks and \(\mathrm{PQI}_N\) comprises \(K{=}7\) narrative checks, yielding \(\mathrm{PQI}\in[0,12]\) (higher is better). We consider a summary \emph{fully compliant} iff \(\mathrm{PQI}(i)=12\), i.e., all checks pass.

\vspace{0.3cm}
\noindent
\textbf{Narrative compliance (\(\mathrm{PQI}_N\))} uses deterministic pattern-based tests to enforce: \textit{(i)} first-person voice, \textit{(ii)} length compliance (90--180 words), \textit{(iii)} grounded opening (name and age appear in the first sentence; country may optionally appear early), \textit{(iv)} single-paragraph prose (no bulleting or list formatting), \textit{(v)} \emph{show-don't-tell} style (no explicit ``I am \(\langle\)trait\(\rangle\)'' self-ascriptions), \textit{(vi)} presence of everyday routine markers, and \textit{(vii)} a grounded, non-moralizing ending.

\vspace{0.3cm}
\noindent
\textbf{Static fidelity (\(\mathrm{PQI}_S\))} uses conservative heuristics to reduce hallucination risk and enforce attribute grounding, including: \textit{(i)} detecting unsupported numerals and suspicious proper nouns not present in the source JSON, \textit{(ii) }preventing explicit leakage of value terminology (e.g., ``WVS'', ``score'', or axis names), and \textit{(iii)} allowing religion mentions only when religion is explicitly provided in the source JSON. Full definitions (regexes, token windows, and lexicons) are provided in the appendix to ensure exact reproducibility.

\vspace{0.3cm}
\noindent
\textbf{PQI Results.}

\textbf{Overall quality is high.} Across \(n=469\) personas, PQI is strongly concentrated near the maximum (\(\mathrm{PQI}\in[0,12]\)), and all summaries are successfully generated (no missing/empty outputs). In total, \(406/469\) personas (\(86.6\%\)) achieve \(\mathrm{PQI}=12\) and \(468/469\) (\(99.8\%\)) score \(\ge 10\) (mean \(=11.85\); median \(=12\)).

\textbf{Component scores are consistently strong.} Static fidelity is near-perfect, with \(\mathrm{PQI}_S=5\) for \(464/469\) personas (\(98.9\%\)). Narrative compliance is also high: \(\mathrm{PQI}_N=7\) for \(410/469\) (\(87.4\%\)) and \(\mathrm{PQI}_N=6\) for \(57/469\) (\(12.2\%\)), suggesting that non-max PQI values mainly reflect minor narrative deviations rather than grounding issues. In addition to PQI, we emit conservative, deterministic audit flags: \(64/469\) personas (\(13.6\%\)) trigger at least one flag, while individual triggers are rare (unsupported-content: \(4/469\), \(0.9\%\); first-person: \(2/469\), \(0.4\%\); and length, religion-policy, and value-leakage: \(1/469\), \(0.2\%\) each). Overall, these results suggest the pipeline produces grounded, stylistically compliant persona summaries at scale, with most flags driven by audit-oriented conservatism rather than systematic issues.

\begin{figure*}[htbp]
\centering
\scriptsize
\forestset{
  menaTree/.style={
    for tree={
      grow'=east,
      parent anchor=east,
      child anchor=west,
      edge={-, line width=0.35pt},
      edge path={
        \noexpand\path[\forestoption{edge}]
          (!u.parent anchor) |- (.child anchor) \forestoption{edge label};
      },
      rounded corners=2pt,
      draw,
      align=left,
      inner sep=1.6pt,
      s sep=1.9mm,
      l sep=5.0mm,
    },
  },
  rootBox/.style={
    fill=gray!12,
    font=\bfseries,
    minimum width=0.65cm,   %
    minimum height=3.8cm,
    align=center,
    inner sep=2pt
  },
  groupRotBox/.style={
    font=\bfseries,
    text width=0.65cm,
    minimum height=2.6cm,
    align=center,
    inner sep=2pt
  },
  leafBox/.style={
    text width=0.80\textwidth,
    inner sep=2pt
  },
  serviceGroup/.style={fill=blue!12},
  serviceLeaf/.style ={fill=blue!4},
  topicGroup/.style  ={fill=green!12},
  topicLeaf/.style   ={fill=green!4}
}

\begin{forest}
  menaTree
  [{\rotatebox{90}{\textbf{Taxonomy}}}, rootBox
    [{\rotatebox{90}{\textbf{Task/Service}}}, groupRotBox, serviceGroup
      [{\textbf{Financial Services}\\ Banking \& Accounts, Payments \& Billing, Credit Cards, Loans \& Mortgages,\\ Insurance (Policies \& Claims)}, leafBox, serviceLeaf]
      [{\textbf{Retail \& Commerce}\\ E-commerce Shopping, Order Tracking \& Returns, Food Delivery \& Takeout, Grocery \& Essentials, \\ Subscriptions \& Memberships}, leafBox, serviceLeaf]
      [{\textbf{Travel \& Mobility}\\ Flight Booking \& Check-in, Hotels \& Accommodation, Car Rental, Ride-hailing \& Taxi,\\Public Transit (Train/Bus)}, leafBox, serviceLeaf]
      [{\textbf{Health \& Wellness}\\ Primary Care \& Clinic Scheduling, Telemedicine \& Symptom Triage, Pharmacy \& Medication Guidance,\\ Mental Health Support, Fitness \& Nutrition Coaching}, leafBox, serviceLeaf]
      [{\textbf{Public Sector \& Utilities}\\ Government Citizen Services, Immigration \& Visa Information, Utilities (Electricity/Water/Gas) \\Support, Telecom \& Internet Support, Emergency \& Disaster Information}, leafBox, serviceLeaf]
      [{\textbf{Productivity, Media \& Education}\\ Calendar \& Scheduling, Email \& Messaging Assistance, IT Helpdesk \& Tech Support,\\ Entertainment \& Streaming Support, Education \& Tutoring}, leafBox, serviceLeaf]
    ]
    [{\rotatebox{90}{\textbf{Knowledge/Topic}}}, groupRotBox, topicGroup
      [{\textbf{Culture \& Society}\\ People \& Everyday Life, Social Interaction, Community \& Civic Life, Occupations \& Careers, Language,\\ Literature, Names, Modern Culture \& Trends, Traditions \& Customs}, leafBox, topicLeaf]
      [{\textbf{Food \& Drink}\\ Traditional \& Regional Cuisines, Cooking \& Eating Customs, Beverages}, leafBox, topicLeaf]
      [{\textbf{Travel, Transport \& Places}\\ Travel, Air Travel, Public Transport, Water Transport, Geography \& Cultural Regions, \\ Famous Landmarks, Architecture \& Design}, leafBox, topicLeaf]
      [{\textbf{History \& Heritage}\\ Heritage \& History, Historical Narratives, National Symbols \& Flags}, leafBox, topicLeaf]
      [{\textbf{Religion \& Spirituality}\\ Religion, Holy Texts \& Scriptures, Places of Worship, Spiritual Practices, Rituals \& Ceremonies}, leafBox, topicLeaf]
      [{\textbf{Arts, Media \& Entertainment}\\ Music, Film \& Animation, Traditional Arts \& Crafts, eSports \& Gaming}, leafBox, topicLeaf]
      [{\textbf{Sports \& Recreation}\\ Sports, Traditional Sports, Outdoor Activities}, leafBox, topicLeaf]
      [{\textbf{Nature \& Environment}\\ Animals, Plants, Weather}, leafBox, topicLeaf]
      [{\textbf{Events \& Celebrations}\\ Festivals \& Celebrations}, leafBox, topicLeaf]
      [{\textbf{Business \& Economy}\\ Business}, leafBox, topicLeaf]
      [{\textbf{Education \& Learning}\\ Education}, leafBox, topicLeaf]
      [{\textbf{Immigration \& Migration}\\ Immigration}, leafBox, topicLeaf]
      [{\textbf{Consumer \& Lifestyle}\\ Clothing \& Fashion, Everyday Objects}, leafBox, topicLeaf]
      [{\textbf{General}\\ General}, leafBox, topicLeaf]
      [{\textbf{Health \& Well-being (Topics)}\\ Healthcare \& Well-being}, leafBox, topicLeaf]
    ]
  ]
\end{forest}

\caption{Root-left taxonomy with two highlighted branches: task/service domains (blue) and knowledge/topic domains (green).}
\label{fig:domain_taxonomy_root_left}
\end{figure*}

\subsection{Persona--Scenario Matching}
\label{persona_scenario_matching}
We hypothesize that personas are most plausible in different domain scenarios. To operationalize this, we automatically align each persona with the most relevant domain scenario(s) by comparing the persona summary text to each scenario text and scoring their similarity. We use a hybrid similarity because embedding-based similarity is robust to paraphrase (capturing semantic relatedness even when surface wording differs), while token overlap anchors the match when domain-specific terms must appear explicitly. This combination produces more robust and realistic persona-scenario pairings.

\noindent
\textbf{Hybrid similarity.}
Let $p$ denote a persona summary and $d$ denote a domain scenario text. We compute \textit{(i)} embedding cosine similarity and \textit{(ii)} token-overlap Jaccard similarity, then combine them with a weighted average:
\begin{align}
s_{\mathrm{emb}}(p,d) &= \cos\!\left(\mathbf{e}_p,\mathbf{e}_d\right), \\
\mathbf{e}_p &= \mathrm{Embed}(p), \quad \mathbf{e}_d = \mathrm{Embed}(d), \\[4pt]
s_{\mathrm{jac}}(p,d) &= \frac{\lvert T(p)\cap T(d)\rvert}{\lvert T(p)\cup T(d)\rvert}, \\[6pt]
s_{\mathrm{hyb}}(p,d) &= w \, s_{\mathrm{emb}}(p,d) + (1-w)\, s_{\mathrm{jac}}(p,d).
\end{align}
We set $w=0.5$ (equal weighting of embedding and token-overlap signals). Finally, we retain only sufficiently similar pairs using a minimum similarity threshold $\tau$:
\begin{equation}
\text{keep}(p,d) \iff s_{\mathrm{hyb}}(p,d) \ge \tau,
\end{equation}
with $\tau = 0.05$ in our experiments.

\begin{table*}[H]
\centering
\vspace{2mm}
\begin{lstlisting}[style=aclprompt]
# Evaluation Prompts
SYSTEM_PROMPT_EVALUATION = """
You are an expert evaluator assessing the quality of AI-generated answers.
Your task is to compare an assistant's answer to a reference answer and rate its quality.

Consider:
- Accuracy: How correct is the information?
- Completeness: Does it address the question fully?
- Relevance: Is the answer on-topic?

Rate on a scale of 1-10 where:
1-3: Poor (incorrect or irrelevant)
4-6: Fair (partially correct but incomplete)
7-8: Good (mostly correct and relevant)
9-10: Excellent (accurate, complete, and relevant)

Provide your rating in this format:
Rating: [[X]]

Where X is an integer from 1 to 10.
"""

USER_PROMPT_ANSWER_EVALUATION = """
Evaluate the following:

Question: {question}

Reference Answer: {reference_answer}

Assistant Answer: {generated_answer}

Provide your rating:
Rating: [[X]]
"""
\end{lstlisting}
\caption{System and User prompts for LLM-based evaluation of model-generated answers.}
\label{tab_evaluation_prompts}
\vspace{-1mm}
\end{table*}

\subsection{Conversation Flow Quality}
To assess the quality of the conversation flow, we asked human annotators to manually rate the generated dialogues after recording the test set. Annotators evaluated coherence, relevance between turns, logical consistency, and overall naturalness using a 5-point Likert scale (annotation guidelines are provided in \ref{annotation_guidelines}). Across 300 conversations, the average rating was 4.9/5, indicating that the generated conversations were generally perceived as coherent and natural.

\begin{figure*}
    \centering
    \includegraphics[width=0.9\linewidth]{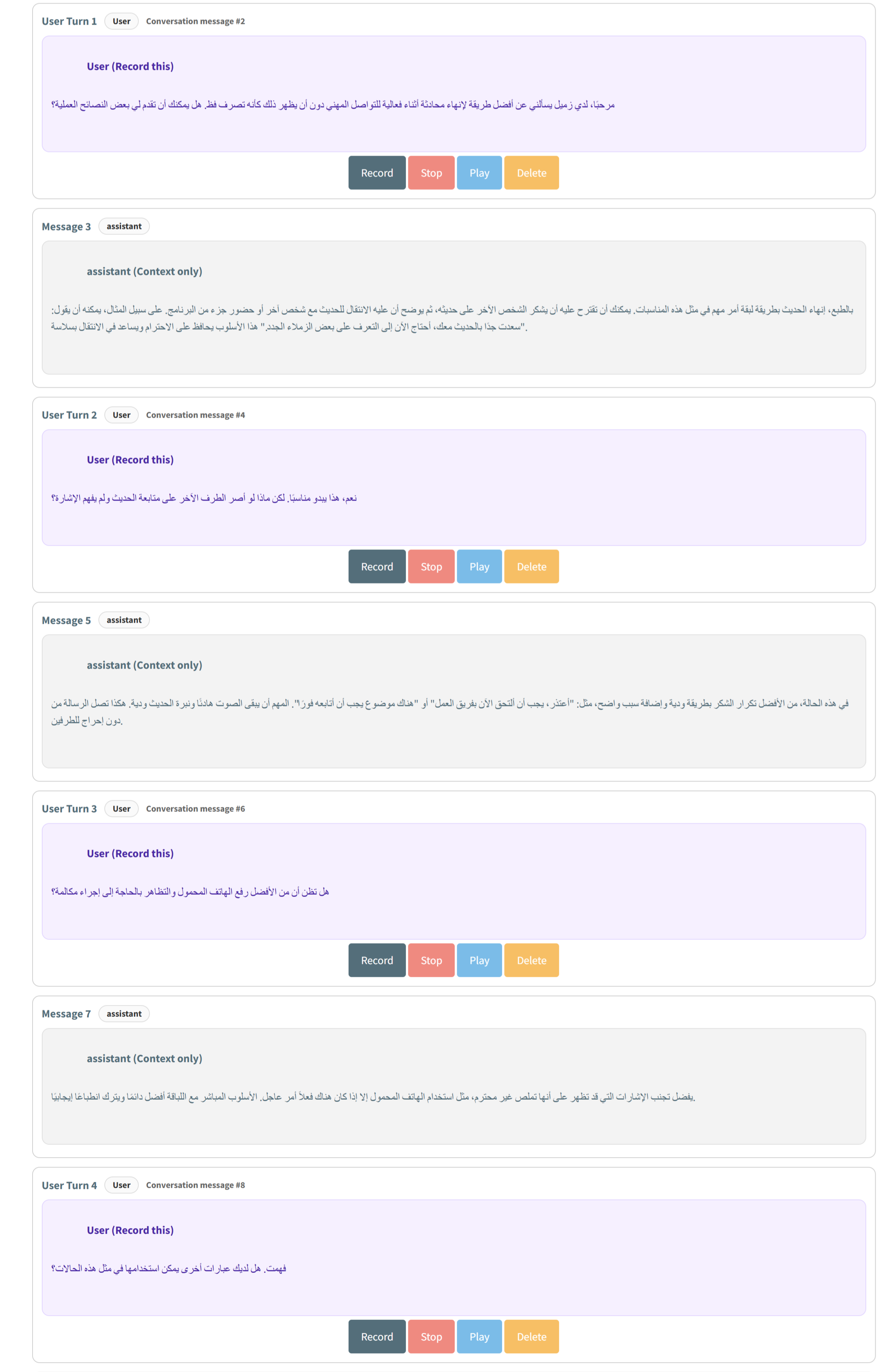}
    \caption{Screenshot of the annotation platform used for conversation recording.}
    \label{fig:placeholder}
\end{figure*}

\section{Data Statistics}\label{appendix_stats}
Table~\ref{tab:speakers_age_group} summarizes the distribution of speakers across different age groups in the \msb{} dataset, highlighting the diversity of the collected speakers.

\begin{figure}[H]
    \centering
    \includegraphics[width=0.99\linewidth]{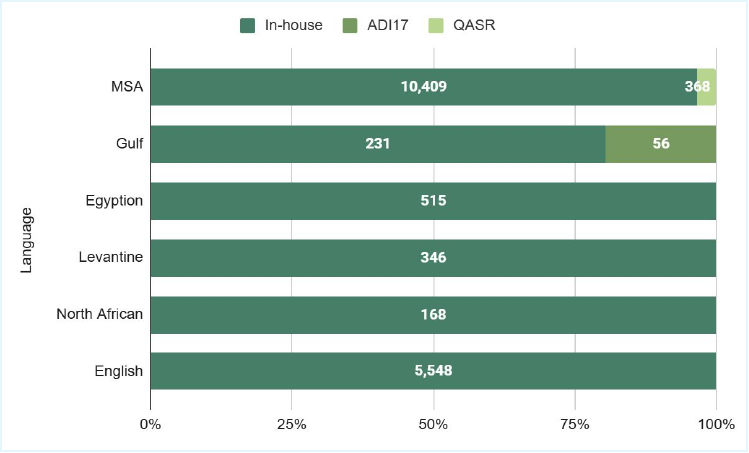}
    \vspace{-0.2cm}
    \caption{Number of speech utterances in the \msb{} dataset, grouped by language or dialect and by dataset origin.}
    \label{fig:dataset_utter_freq}
\end{figure}

Figure~\ref{fig:dataset_utter_freq} illustrates the distribution of speech utterances across languages or dialects, further grouped by dataset origin to show the contribution of each source to the overall dataset.
\begin{table}[H]
\centering
\small
\setlength{\tabcolsep}{4pt}
\begin{tabular}{lcccc}
\toprule
\textbf{Age group} &
\textbf{\# Speakers} \\
\midrule
<20 & 4\\
20-29 & 50\\
30-39 & 19\\
40-49 & 20\\
50-59 & 16\\
60+ & 12 \\
\bottomrule
\end{tabular}
\caption{Distribution of speakers across age groups in the \msb{} dataset.}
\label{tab:speakers_age_group}
\vspace{-0.3cm}
\end{table}

\section{Data Recording}
We hired annotators for recording through a third-party company and compensated at the standard hourly rate that is reasonable for their location. All annotators were professionals fluent in both Arabic and English and held at least a bachelor's degree. Each annotator signed a non-disclosure agreement (NDA) specifying all permitted uses of the data.

\subsection{Annotation Guidelines}
\label{annotation_guidelines}
Voice Recording Instructions
\begin{enumerate}
    \item \textbf{Review the conversation: }For each turn, you will see an Assistant message (context) and a User message (the text you must record).
    \item \textbf{Record each User turn}: Click Record, read the User text aloud exactly as written, then click Stop.
    \item \textbf{Playback:} Click Play to listen and confirm your audio is clear and complete.
    \item \textbf{Re-record if needed:} If you made a mistake or the audio is noisy/cut off, click Delete and record again.
    \item Check your recording: Always press Play. If the audio is too quiet, clipped, or incorrect, Delete and re-record. After recording all 3 user turns, select a conversation-flow rating from 1 (Bad) to 5 (Good).
    \item \textbf{Submit:} You can submit only after all 3 recordings are present and a rating is selected.
\end{enumerate}

\subsection{Instructions}
\textbf{What to record:} Record only the text shown in the User (Record this) box. The Assistant text is provided for context only and should not be recorded.

Recording rules
\begin{itemize}
    \item Read exactly: Do not paraphrase, translate, correct grammar, or add extra words. Read the user text word-for-word.
    \item Speak naturally and clearly: Use a normal speaking pace and clear pronunciation.
    \item Follow punctuation: Questions should sound like questions. Pause naturally at commas and sentence boundaries.
    \item One speaker, low noise: Record in a quiet place. Avoid TV/music, loud fans, or other people speaking.
    \item Check your recording: Always press Play. If the audio is too quiet, clipped, or incorrect, Delete and re-record.
\end{itemize}

\textbf{Conversation Flow Quality (1–5)}
After you finish recording all 3 turns, rate the overall conversation flow for the snippet shown (assistant + user turns). Flow quality reflects how natural and coherent the back-and-forth is.

\textbf{Rate based on:} relevance between turns, logical coherence, consistency, and naturalness of the dialogue.

\textbf{Do NOT rate based on:} whether you agree with the content, factual correctness, or your own recording quality.

\begin{itemize}[noitemsep,topsep=0pt,leftmargin=*,labelsep=.5em]
    \item \textbf{5 (Good):} Very natural and coherent; user replies clearly match the assistant context.
    \item \textbf{4:} Mostly coherent; only minor awkwardness or small mismatch.
    \item \textbf{3:} Mixed; noticeable issues but still understandable overall.
    \item \textbf{2:} Poor; multiple mismatches, unclear transitions, or unnatural dialogue.
    \item \textbf{1 (Bad):} Very incoherent; turns are disconnected, contradictory, or confusing.
\end{itemize}

\textbf{Troubleshooting}
\begin{itemize}[noitemsep,topsep=0pt,leftmargin=*,labelsep=.5em]
    \item If you cannot record, make sure your browser has microphone permission enabled, then refresh the page.
    \item If playback is silent, check your microphone input and re-record closer to the mic.
\end{itemize}

\section{Fine-tuning Parameters}
\label{sec_app_fine_tuning}
LoRa fine-tuning was performed with rank $r = 32$, scaling factor $\alpha = 64$, and dropout $p = 0.05$. Training used a per-device batch size of 2 with gradient accumulation over 16 steps, a learning rate of $3 \times 10^{-5}$, and a warmup ratio of 0.2. The maximum sequence length was set to 8192. All experiments were conducted on 8 $\times$ H100 GPUs.

\begin{table}[!tbh]
\centering
\small
\setlength{\tabcolsep}{3.0pt}

\scalebox{0.98}{%
\begin{tabular}{lcc|cc|cc}
\toprule
\multirow{2}{*}{\textbf{ID}} &
\multirow{2}{*}{\textbf{\#Conv}} &
\multirow{2}{*}{\textbf{\#Turns}} &
\multicolumn{2}{c|}{\textbf{Conv.-level}} &
\multicolumn{2}{c}{\textbf{Turn-level}} \\
\cmidrule(lr){4-5}
\cmidrule(lr){6-7}
& & &
\textbf{APR} & \textbf{ARS} &
\textbf{APR} & \textbf{ARS} \\
\midrule

\multicolumn{7}{c}{\textbf{(a) Gemini-2.5 Pro}} \\
\midrule
0 & 52 & 260 & 0.712 & 0.858 & 0.827 & 0.968 \\
1 & 51 & 254 & 0.824 & 0.910 & 0.823 & 0.973 \\
2 & 49 & 245 & 0.755 & 0.865 & 0.792 & 0.966 \\
3 & 17 & 84  & 0.824 & 0.938 & 0.905 & 0.988 \\
4 & 17 & 85  & 0.882 & 0.976 & 0.965 & 0.995 \\
5 & 8  & 40  & 0.625 & 0.825 & 0.700 & 0.945 \\
6 & 4  & 20  & 1.000 & 1.000 & 0.800 & 0.975 \\
7 & 3  & 15  & 1.000 & 1.000 & 1.000 & 1.000 \\
8 & 1  & 5   & 0.000 & 0.600 & 0.600 & 0.860 \\
9 & 1  & 5   & 1.000 & 1.000 & 1.000 & 1.000 \\

\midrule
\multicolumn{7}{c}{\textbf{(b) GPT-Audio}} \\
\midrule
0 & 52 & 260 & 0.712 & 0.885 & 0.808 & 0.962 \\
1 & 51 & 254 & 0.765 & 0.878 & 0.756 & 0.957 \\
2 & 49 & 245 & 0.735 & 0.861 & 0.771 & 0.960 \\
3 & 17 & 84  & 0.824 & 0.965 & 0.835 & 0.984 \\
4 & 17 & 85  & 0.882 & 0.941 & 0.859 & 0.962 \\
5 & 8  & 40  & 0.750 & 0.875 & 0.800 & 0.968 \\
6 & 4  & 20  & 1.000 & 1.000 & 1.000 & 1.000 \\
7 & 3  & 15  & 1.000 & 1.000 & 1.000 & 1.000 \\
8 & 1  & 5   & 1.000 & 1.000 & 1.000 & 1.000 \\
9 & 1  & 5   & 1.000 & 1.000 & 0.800 & 0.980 \\
\bottomrule
\end{tabular}%
}

\vspace{-0.1cm}
\caption{Speaker-wise results for Gemini-2.5 Pro and GPT-Audio at conversation and turn levels. \#Conv and \#Turns denote the number of conversations and turns per speaker.}
\label{tab:speaker_wise_results}
\vspace{-0.2cm}
\end{table}

\begin{table*}[!tbh]
\centering
\small
\setlength{\tabcolsep}{4pt}
\scalebox{0.9}{
\begin{tabular}{lcc|cccc|cccc}
\toprule
\multirow{2}{*}{\textbf{Topic}} &
\multirow{2}{*}{\textbf{\#Conv}} &
\multirow{2}{*}{\textbf{\#Turns}} &
\multicolumn{4}{c|}{\textbf{Fanar-C-2-27B}} &
\multicolumn{4}{c}{\textbf{Gemini-2.5 pro}} \\
\cmidrule(lr){4-7} \cmidrule(lr){8-11}
& & &
\multicolumn{2}{c}{\textbf{Conv.-level}} &
\multicolumn{2}{c|}{\textbf{Turn-level}} &
\multicolumn{2}{c}{\textbf{Conv.-level}} &
\multicolumn{2}{c}{\textbf{Turn-level}} \\
\cmidrule(lr){4-5} \cmidrule(lr){6-7}
\cmidrule(lr){8-9} \cmidrule(lr){10-11}
& & &
\textbf{APR} & \textbf{ARS} &
\textbf{APR} & \textbf{ARS} &
\textbf{APR} & \textbf{ARS} &
\textbf{APR} & \textbf{ARS} \\
\midrule
Culture \& Society                & 47 & 235 & 0.447 & 0.762 & 0.727 & 0.941 & 0.894 & 0.974 & 0.898 & 0.984 \\
Travel, Transport \& Places      & 24 & 120 & 0.333 & 0.717 & 0.583 & 0.893 & 0.875 & 0.942 & 0.900 & 0.986 \\
Productivity, Media \& Education & 18 & 90  & 0.222 & 0.689 & 0.500 & 0.893 & 0.722 & 0.900 & 0.800 & 0.977 \\
Religion \& Spirituality         & 17 & 85  & 0.471 & 0.812 & 0.753 & 0.936 & 0.941 & 0.988 & 0.965 & 0.996 \\
Health \& Wellness               & 15 & 74  & 0.467 & 0.770 & 0.676 & 0.936 & 0.667 & 0.853 & 0.824 & 0.965 \\
Food \& Drink                    & 13 & 65  & 0.231 & 0.769 & 0.692 & 0.935 & 1.000 & 1.000 & 0.954 & 0.995 \\
Education \& Learning            & 11 & 55  & 0.455 & 0.791 & 0.750 & 0.944 & 0.909 & 0.964 & 0.927 & 0.989 \\
General                          & 9  & 45  & 0.333 & 0.711 & 0.644 & 0.904 & 0.667 & 0.844 & 0.844 & 0.967 \\
History \& Heritage              & 8  & 40  & 0.375 & 0.750 & 0.750 & 0.938 & 0.875 & 0.975 & 0.975 & 0.993 \\
Arts, Media \& Entertainment     & 7  & 35  & 0.286 & 0.514 & 0.429 & 0.849 & 1.000 & 1.000 & 0.829 & 0.983 \\
Events \& Celebrations           & 5  & 25  & 0.800 & 0.840 & 0.640 & 0.964 & 0.600 & 0.880 & 0.880 & 0.984 \\
Sports \& Recreation             & 5  & 25  & 0.000 & 0.440 & 0.440 & 0.824 & 0.800 & 0.960 & 0.840 & 0.968 \\
Travel \& Mobility               & 5  & 25  & 0.750 & 0.950 & 0.550 & 0.910 & 0.200 & 0.520 & 0.440 & 0.940 \\
Consumer \& Lifestyle            & 4  & 20  & 0.250 & 0.650 & 0.600 & 0.925 & 0.750 & 0.950 & 0.850 & 0.960 \\
Nature \& Environment            & 4  & 20  & 0.000 & 0.650 & 0.650 & 0.940 & 0.750 & 0.950 & 0.950 & 0.990 \\
Public Sector \& Utilities       & 4  & 19  & 0.250 & 0.500 & 0.450 & 0.900 & 0.250 & 0.650 & 0.650 & 0.940 \\
Retail \& Commerce               & 4  & 20  & 0.000 & 0.600 & 0.200 & 0.865 & 0.250 & 0.850 & 0.800 & 0.945 \\
Business \& Economy              & 3  & 15  & 0.667 & 0.933 & 0.933 & 0.987 & 1.000 & 1.000 & 0.867 & 0.987 \\
Health \& Well-being (Topics)    & 3  & 15  & 0.333 & 0.800 & 0.800 & 0.947 & 0.667 & 0.733 & 0.733 & 0.947 \\
\bottomrule
\end{tabular}
}
\vspace{-0.2cm}
\caption{Topic-wise evaluation results for Gemini-2.5 Pro, on human audio, and Fanar-C-2-27B, on Fanar (Aura)-transcribed human audio. Reported at both conversation and turn levels using APR and ARS metrics.}
\label{tab:gemini_topic_results}
\vspace{-0.3cm}
\end{table*}

\section{Additional Experimental Results}
\label{sec_app_exp_results}

\paragraph{Statistical Significance.} To assess whether differences between systems are reliable, we apply paired significance tests over 206 conversations using three methods: paired bootstrap (10,000 resamples), Wilcoxon signed-rank test on ARS, and McNemar's exact test on APR. All tests use the conversation as the resampling unit since turns within a conversation are not independent. We compare 64 system pairs across three conditions: models on generated speech, models on human recordings, and generated vs. human speech for the same model.

Of the 64 pairs, 54 differ significantly at p<0.05, with most at $p<0.001$. The 10 non-significant pairs %
involve top-performing systems (e.g., Gemini vs. GPT-Audio on generated speech; Gemini vs. ASR+GPT on human speech), indicating comparable performance among the strongest models. For 5 of 8 models, the difference between generated and human speech is not significant, supporting the validity of using synthesized speech in the benchmark.

\subsection{Speaker-wise Experimental Results}
Table~\ref{tab:speaker_wise_results} presents the speaker-wise evaluation results for recorded conversations accross audio models (GPT-audio and Gemini). We report both conversation and turn level APR and ARS scores. Performance was stable across speakers with sufficient data (more than 17 conversations) but varied for those with limited data. Overall, there is no clear indication of systematic bias toward specific speakers.
GPT-audio showed greater robustness to speaker variability, particularly for underrepresented speakers. However, speaker 5 showed lower scores on both models, suggesting potential challenges related to audio quality or speech patterns.

\subsection{Topic-wise Experimental Results}
Table  \ref{tab:gemini_topic_results} presents topic-wise evaluation results for Gemini-2.5 pro, the best-performing audio model, on human audio and Fanar-C-2-27B, the best-performing Arabic-centric model on transcibed text of human audio. Fanar-C-2-27B achieves its best performance on structured, task-oriented topics such as Events \& Celebrations (APR: 0.80) and Travel \& Mobility (APR: 0.75), while failing on descriptive domains like Sports \& Recreation and Nature \& Environment (APR: 0.00).
Gemini-2.5 pro excels on conversationally rich topics including Food \& Drink, Arts \& Entertainment, and Business \& Economy (APR: 1.00), but underperforms on service-oriented topics such as Travel \& Mobility and Public Sector \& Utilities (APR: 0.20-0.25).
Both models share a common weakness on Travel \& Mobility and Retail \& Commerce, suggesting these domains present inherent challenges for conversational evaluation regardless of model architecture.

\subsection{Gender-wise Experimental Results}
Table \ref{tab:gender_wise_results} presents gender-wise evaluation results. For synthetic audio, \textit{gender} refers to the reference audio used for voice cloning, which matches the assigned persona gender. For human recordings, \textit{gender} refers to the speaker's self-identified gender.

Both models exhibit higher performance on female voices across synthetic and human audio inputs. The performance gap is most pronounced for Gemini on human recordings (APR: 0.79 female vs 0.58 male), while GPT-Audio demonstrates more balanced performance across genders (APR: 0.83 female vs 0.73 male). This disparity may be influenced by sample imbalance (143 female vs 60 male conversations in human recordings) or differences in acoustic characteristics between genders.

\begin{table}[t]
\centering
\scriptsize
\setlength{\tabcolsep}{2.3pt}

\scalebox{0.98}{%
\begin{tabular}{llcc|cc|cc}
\toprule
\multirow{2}{*}{\textbf{Audio}} &
\multirow{2}{*}{\textbf{Gender}} &
\multirow{2}{*}{\textbf{\#Conv}} &
\multirow{2}{*}{\textbf{\#Turns}} &
\multicolumn{2}{c|}{\textbf{Conv.-level}} &
\multicolumn{2}{c}{\textbf{Turn-level}} \\
\cmidrule(lr){5-6}
\cmidrule(lr){7-8}
& & & &
\textbf{APR} & \textbf{ARS} &
\textbf{APR} & \textbf{ARS} \\
\midrule

\multicolumn{8}{c}{\textbf{(a) Gemini-2.5 Pro}} \\
\midrule
\multirow{2}{*}{Synthetic}
& Female & 83  & 415 & 0.807 & 0.896 & 0.786 & 0.968 \\
& Male   & 123 & 614 & 0.740 & 0.891 & 0.811 & 0.960 \\
\addlinespace[1pt]
\multirow{2}{*}{Human}
& Female & 143 & 715 & 0.790 & 0.922 & 0.853 & 0.973 \\
& Male   & 60  & 299 & 0.583 & 0.830 & 0.742 & 0.946 \\

\midrule
\multicolumn{8}{c}{\textbf{(b) GPT-Audio}} \\
\midrule
\multirow{2}{*}{Synthetic}
& Female & 83  & 415 & 0.795 & 0.901 & 0.836 & 0.970 \\
& Male   & 123 & 614 & 0.764 & 0.886 & 0.830 & 0.972 \\
\addlinespace[1pt]
\multirow{2}{*}{Human}
& Female & 143 & 715 & 0.832 & 0.943 & 0.892 & 0.985 \\
& Male   & 60  & 299 & 0.733 & 0.883 & 0.813 & 0.967 \\
\bottomrule
\end{tabular}%
}

\vspace{-0.1cm}
\caption{Gender-wise results for Gemini-2.5 Pro and GPT-Audio on synthetic and human speech, evaluated at conversation and turn levels.}
\label{tab:gender_wise_results}
\vspace{-0.2cm}
\end{table}

\section{Prompts}

\subsection{Prompt Details}
\label{app:persona-prompts}

Figure~\ref{app_prompt_for_summary} presents the system and user prompts used for persona-to-summary generation.
Table~\ref{tab:app_topic_prompt} shows the prompt used to generate domain-specific conversation topics for each taxonomy path.
Table~\ref{tab_app_scenario_prompt} provides the prompt used to generate user-centric conversational scenarios for each topic.
Tables~\ref{tab_app_dialogue_prompts_1} and~\ref{tab_app_dialogue_prompts_2} present the system and user prompts used for dialogue generation with strict JSON output and language constraints.
\begin{figure*}[htbp]
\centering
\begin{minipage}{0.98\textwidth}
\vspace{0.5em}
\noindent\textbf{System Prompt.}
\begin{lstlisting}[style=promptstyle]
You are a specialized AI "Persona Profiler." Your sole task is to take a single JSON object containing persona attributes-aligned with a WVS-inspired persona schema---and produce a coherent, natural, first-person persona summary.

Input: JSON object with persona attributes (schema below).

Output: strict JSON only:
{"summary_first_person":"<120--180 words>"}

Hard constraints:
- One paragraph. First-person only ("I..."). Natural, grounded tone.
- 90--180 words.
- Opening sentence MUST include: persona_name + speaker_age + (city if present) + speaker_nationality.
- Use ONLY explicit JSON facts. No invention, inference, stereotypes, or extra backstory.
- Do not output IDs (speaker_id, script_id, region) unless user explicitly asks.

Content selection (use if present):
- Identity/core: persona_name, speaker_age, gender (only if it fits naturally), speaker_mother_tongue, native, city, speaker_nationality.
- Life context: Use marital_status and household_type to describe living arrangements and family situation, ensuring they are mutually consistent.
- Work/education: Use education_level and profession together, ensuring the stated role is plausible and consistent with the education level.
- Digital/AI: Reflect technology access and AI familiarity using digital_access.device, digital_access.connectivity, and digital_access.ai_competence_level; also consider life context, education, and profession.
- AI Use Cases: Integrate them naturally based on the ai_use_case information and other personal attributes. Do not assume enthusiasm for AI---reflect only what the fields support.
- Religion: Mention only if religion exists; keep neutral.
- Include small, ordinary, concrete details (routines, habits, regular activities).
- Interests, activities and hobbies must appear organically within the flow of daily life and they should be consistent with age.

Values/profiles:
- wvs_profile: NEVER mention numbers, axes names, "WVS", or "scores". If used at all, reflect implicitly via everyday preferences/actions. If a value is null, ignore it. Do not "explain" or psychoanalyze.
- ocean: NEVER mention trait names or numbers. Do not write "I am X" traits. If used, show via behavior (e.g., planning habits) without labeling.

Negative/contradictions:
- If conflicts/negatives exist, preserve them plainly; no "growth/lesson" framing.

Ending:
- No abstract/moral reflections ("I've learned...", "In the end..."). End with a routine, near-term goal, or ongoing interest.

Failure rule:
- If you cannot satisfy opening/length/JSON-only, output {"summary_first_person":""}.

Internal self-check before final:
[ ] first-person; [ ] 120--180 words; [ ] opening fields present; [ ] no IDs; [ ] no invented facts;
[ ] no WVS/OCEAN labels or numbers; [ ] no "I am <trait>"; [ ] grounded ending; [ ] strict JSON only.
\end{lstlisting}

\noindent\textbf{User Prompt Template.}
\begin{lstlisting}[style=promptstyle]
Generate a first-person persona summary from the following JSON data consisting of persona attributes.

Input JSON:
{json_object}
\end{lstlisting}

\end{minipage}
\caption{System and user prompts used for persona-to-summary generation.}
\label{app_prompt_for_summary}
\end{figure*}

\begin{table*}[htbp]
\centering
\vspace{2mm}
\begin{lstlisting}[style=aclprompt]
prompt = """You are an expert in conversational AI.
Domain: "{domain_path}"

Task:
Generate exactly {num_topics} specific, relevant conversation topics for this domain.

Requirements:
1) Topics must be specific and actionable user-intent areas.
2) Topics must represent common user intents and conversation flows.
3) Topics must be diverse and cover different aspects of the domain (avoid near-duplicates).
4) Use clear, concise topic names (2--6 words each).
5) Prefer topics commonly relevant in MENA contexts when applicable. Do NOT invent country-specific facts unless implied.

Output format (STRICT):
Return ONLY valid JSON (no markdown, no explanations).
The JSON must be an object with a single key "topics" whose value is an array of strings.
Example:
{{"topics": ["Topic 1", "Topic 2"]}}

Generate exactly {num_topics} topics.
"""
\end{lstlisting}
\vspace{-1mm}
\caption{Prompt used to generate domain-specific conversation topics for each taxonomy path.}
\label{tab:app_topic_prompt}
\end{table*}

\begin{table*}[htbp]
\centering
\vspace{2mm}
\begin{lstlisting}[style=aclprompt]
PROMPT = """You are an expert in conversational AI and user experience design.
Given:
- Domain path: "{domain_path}"
- Topic: "{topic}"

Generate {num_scenarios} realistic conversational scenarios that users might have when interacting with a system for this topic.

Each scenario should be:
1. A natural, user-focused description of what the user wants to accomplish
2. Specific and actionable (not vague)
3. Representative of real-world user intents
4. Written in a conversational, user-centric way (e.g., "A customer wants to...", "A parent needs to...")

Return ONLY a JSON array of scenario strings, no explanations, no markdown.
"""
\end{lstlisting}
\caption{Prompt used to generate user-centric conversational scenarios for each topic.}
\label{tab_app_scenario_prompt}
\vspace{-1mm}
\end{table*}

\begin{table*}[htbp]
\centering
\vspace{2mm}
\begin{lstlisting}[style=aclprompt]
SYSTEM_PROMPT = """
You are a helpful assistant.

OUTPUT FORMAT (STRICT)
- Return exactly one JSON object with a "messages" array.
- Each item in "messages" must be an object with:
  - "role": "user" or "assistant"
  - "content": message text
- No extra text, no markdown, no code fences.

INITIATION (VARY THIS)
- Vary who starts the conversation:
  - ~50-60%
  - ~40-50%

LANGUAGE (CRITICAL)
- All user messages MUST be in PURE {language}.
- No English words, no code-switching, no transliterated English.
- If {language} is Arabic, use ONLY Modern Standard Arabic.
  - Do NOT use dialect words/particles.
  - If any dialect word appears, rewrite that message into MSA before output.
- If the assistant starts (outbound), its first message MUST also be in PURE {language} (and in MSA if Arabic).

VALID JSON EXAMPLE:
{{"messages":[{{"role":"user","content":"..."}},{{"role":"assistant","content":"..."}}]}}
"""


DIALOGUE_GENERATION_PROMPT = """
You are an expert in conversational AI and natural language generation.
Your task is to generate a realistic, natural dialogue between a user and an AI assistant.

USER INFORMATION (Persona):
- Name: {user_name}
- Age: {user_age}
- Nationality: {user_nationality}
- Language: {language}  (If {language} is Arabic, it MUST be Modern Standard Arabic / MSA)
- Gender: {user_gender}
- City: {user_city}
- Education Level: {user_education_level}
- Profession: {user_profession}
- Marital Status: {user_marital_status}
- Household Type: {user_household_type}
- Religion: {user_religion}
- Digital Access: {user_digital_access}
- AI Use Cases: {user_ai_use_cases}
- Values Profile: {user_wvs_profile}
- Persona Summary: {summary}

TOPIC INFORMATION:
- Domain: {domain}
- Topic: {topic}
- Scenario: {scenario}
- Max Messages: {max_messages}

\end{lstlisting}
\caption{System and user prompts for dialogue generation with JSON output and language constraints. (Part 1)}
\label{tab_app_dialogue_prompts_1}
\vspace{-1mm}
\end{table*}

\begin{table*}[htbp]
\centering
\small
\vspace{2mm}
\begin{lstlisting}[style=aclprompt]
CRITICAL REQUIREMENTS:
1. Generate a natural, realistic SPOKEN conversation between a user and an AI assistant.
2. The user should speak in the target language and reflect their persona characteristics.

3. CRITICAL LANGUAGE REQUIREMENTS:
   - The user MUST speak in pure {language} without mixing English words, code-switching, transliterations, or dialectal variations.
   - Use ONLY native words from {language}.
   - Replace English technical terms with native equivalents.
     - Arabic examples (MSA): use `\<هاتف / هاتف محمول>` instead of transliterations like `\<موبايل>`.
     - Use `\<على الإنترنت>` sparingly and prefer native descriptive phrases when possible.
   - Avoid transliterated English words; use native equivalents.
   - Example (Arabic): instead of `\<آب / أب>` for app use `\<تطبيق>` instead of `\<بلاتفورم>` use `\<منصة/منصة إلكترونية>` or any appropriate Arabic description.

4. ARABIC MODE (STRICT) - Applies ONLY if {language} is Arabic / MSA:
   - Use ONLY Modern Standard Arabic `\<(العربية الفصحى الحديثة)>` for BOTH user and assistant.
   - Do NOT use dialect particles/negations/words such as:
   `\<(إيش، شو، شنو، وين، ليش، مو، مش، عايز، بدّي، حابب، كده، كذا، هلا، هلّق، رح، لسا، بزاف، برشا).>`

   - Spoken MSA style: keep sentences short and natural for speech, but still MSA.
     - Allowed MSA fillers: `\<(مم…، حسنًا…، دعني أفهم…، لحظة…، على مهلك…).>`

   - SELF-CHECK: If any dialect word appears, rewrite that message into MSA before output.

5. SPOKEN CONVERSATION STYLE:
   - The dialogue must sound like real spoken conversation, not written text.
   - Use short, direct messages, clarifications, and back-and-forth.
   - Avoid overly formal or written-style language.
   - The conversation should feel authentic and not scripted.

6. INITIATION (VARY THIS):
   - Sometimes the USER initiates (about 50-60%
   - Sometimes the ASSISTANT initiates (about 40-50%
   - If assistant starts, the greeting and initial message MUST be in {language}.
   - If {language} is Arabic/MSA, the greeting MUST be in MSA (not dialect).

OUTPUT FORMAT (STRICT):
Return ONLY a JSON object with a "messages" key containing an array of message objects.
Each message object must have:
- "role": "user" or "assistant"
- "content": message text

Inbound example:
{{"messages":[{{"role":"user","content":"Message in {language}"}},{{"role":"assistant","content":"Response"}}, ...]}}

Outbound example:
{{"messages":[{{"role":"assistant","content":"Greeting in {language}"}},{{"role":"user","content":"Response in {language}"}}, ...]}}

IMPORTANT:
- First message can be from either "user" or "assistant" (vary this).
- User must always use PURE {language} with no English mixing.
- If {language} is Arabic/MSA, BOTH sides must be strictly MSA with no dialect words.
- Generate a realistic SPOKEN conversation with natural flow. Maximum {max_messages} messages.
"""
\end{lstlisting}
\caption{System and user prompts for dialogue generation with JSON output and language constraints. (Part 2)}
\label{tab_app_dialogue_prompts_2}
\vspace{-1cm}
\end{table*}

\section{Data and Resource Release}
\label{sec:data_release}
We will release the supporting code, documentation, and dataset under the CC BY-NC-SA 4.0 license to facilitate adoption, reproducibility, and future extensions.

\end{document}